\documentclass[letterpaper,twocolumn,11pt,accepted=2017-05-09]{quantumarticle}
\pdfoutput=1

\usepackage[numbers]{natbib}
\usepackage{tikz}
\usepackage[strict]{revquantum}
    \newaffil{MSQ}{
        Microsoft Quantum,
        Microsoft,
        Redmond, WA, United States
    }

\usepackage{xcolor}
    \colorlet{comment}{cud-bluish-green!30!black!70}

\usepackage{algpseudocode}

    \renewcommand{\Comment}{\hfill{\footnotesize/\!\!/} }
    \algblockdefx[Arguments]%
        {Arguments}{EndArguments}%
        {
            \color{comment}
            {\footnotesize \(\blacksquare\) \emph{Arguments}} \normalsize%
        }%
        {\color{black}}
    \algsetblock[Arguments]{Arguments}{EndArguments}{}{0.325cm}
    \renewcommand{\inlinecomment}[1]{{\color{comment} \Comment {\footnotesize #1} \normalsize}}
    \renewcommand{\linecomment}[1]{\State {\color{comment} {\footnotesize/\!\!/ #1} \normalsize}}
    \newcommand{\seccomment}[1]{%
        \vskip0.5em%
        \State {\color{comment} \footnotesize \(\blacksquare\) \emph{#1}} \normalsize
    }

    \newrm{inv}
    \newrm{unwind}
    \newrm{ess}
    \newcommand{\nexp}{\ensuremath{n_{\textrm{exp}}}}
    \newcommand{\iexp}{\ensuremath{i_{\textrm{exp}}}}
    \newcommand{\tch}{\ensuremath{\tau_{\textrm{check}}}}
    \algnewcommand\Raise{\textbf{raise} }

    \algnotext{EndFor}
    \algnotext{EndIf}
    \algnotext{EndWhile}
    \algnotext{EndFunction}

\newcommand{\figurefolder}{../fig}

\input{./Qcircuit}

\renewcommand{\figurefolder}{fig}
\begin{document}

\title{Using Random Walks for Iterative Phase Estimation}
\date{authors in alphabetical order}

\author{Cassandra Granade}
    \affilMSQ

\author{Nathan Wiebe}
\affiliation{
    Department of Computer Science, University of Toronto, Toronto, Canada}
\affiliation{Pacific Northwest National Laboratory, Richland, USA
}
\affiliation{
    University of Washington, Department of Physics, Seattle, USA}
\begin{abstract}
    In recent years there has been substantial development in algorithms for quantum phase estimation. In this work we provide a new approach to online Bayesian phase estimation that achieves Heisenberg limited scaling that requires exponentially less classical processing time with the desired error tolerance than existing Bayesian methods.  
    This practically means that we can perform an update in microseconds on a CPU as opposed to milliseconds for existing particle filter methods.  Our approach assumes that the prior distribution is Gaussian and exploits the fact, when optimal experiments are chosen, the mean of the prior distribution is given by the position of a random walker whose moves are dictated by the measurement outcomes.  We then argue from arguments based on the Fisher information that our algorithm provides a near-optimal analysis of the data.  This work shows that online Bayesian inference is practical, efficient and ready for deployment in modern FPGA driven adaptive experiments.
\end{abstract}

\maketitle

\section{Introduction}

Phase estimation is widely used throughout quantum information to learn the eigenvalues of unitary operators.
This is a critical step in realizing computational speedups from algorithms such as Shor's algorithm~\cite{shor1994algorithms}, versions of the linear systems algorithm~\cite{harrow2009quantum} and quantum simulation~\cite{reiher2017elucidating,abrams1997simulation,von2021quantum,su2021fault}.
Moreover, the eigenvalues of unitary operators can carry information about physical parameters, such that phase estimation is critical in quantum metrology applications as well~\cite{berry_optimal_2001,hentschel_machine_2010} and similar ideas can be used in Hamiltonian learning protocols as well~\cite{wiebe_hamiltonian_2014,zintchenko_randomized_2016,wang2017experimental}.

One very popular approach to minimizing the quantum resources required for phase estimation is \emph{iterative phase estimation}~\cite{kitaev2002classical,svore_faster_2013, kimmel2015robust}, in which one repeatedly uses the unitary of interest to write a phase onto a single-qubit register, and then processes the data collected from these measurements using classical resources.
The data obtained from each iteration can then be used to determine the next measurement to perform, such that classically adaptive protocols are naturally expressed in the iterative phase estimation framework.
Across applications for phase estimation, there is a significant benefit to formulating approaches to iterative phase estimation that allow for adaptivity and data processing with very modest classical resources.
For instance, \citet{wiebe_efficient_2015} have introduced \emph{rejection filtering}, an iterative phase estimation algorithm that is modest enough to be compatible with modern experimental control hardware such as field-programmable gate arrays (FPGAs).
Recent experimental demonstrations have shown the effectiveness of rejection filtering in modern applications \cite{paesani_experimental_2017}.

In this work, we simplify the classical processing requirements still further, providing a new algorithm that uses approximately 250 bits of classical memory between iterations and that is entirely deterministic conditioned on the experimental data record.
Our new algorithm is thus uniquely well-suited not only for implementation on FPGAs, but as an application-specific integrated circuit (ASIC) or on tightly memory-bound microcontroller platforms, which are common in certain cryogenic applications where it is desirable to have a small low-power controller located as close as possible to the base plate of the dilution refrigerator for latency purposes despite the limited cooling power available at such stages.
The ability to use such simple control hardware in turn significantly reduces the cost required for implementing quantum algorithms~and new metrology protocols such as distributed sensors \cite{eldredge_optimal_2016} in near term quantum hardware.

\section{Review of Bayesian Phase Estimation}

Concretely, consider a family of unitary operators $U(t)$ for a real parameter $t$ and a quantum state $\ket{\omega}$ such that for all $t$ $U(t)\ket{\omega} = \e^{\ii \omega t}\ket{\omega}$ for an unknown real number $\omega$.
We are then interested in performing controlled applications of $U(t)$ on a copy of the state $\ket{\omega}$ in order to learn $\omega$.

There are many ways that the learning problem for phase estimation can be formalized.
While the most commonly used methods have historically estimated the phase in a bit-by-bit fashion \cite{kitaevay_quantummeasurementsabelian_1995}, Bayesian approaches to learning $\omega$ have recently gained in popularity because of their robustness and their statistical efficiency \cite{wiebe_efficient_2015}.
The idea behind such methods is to quantify the uncertainty in $\omega$ via a prior probability distribution, $\Pr(\omega)$.
Then conditioned on measuring an outcome $d$, the probability distribution describing the uncertainty in $\omega$ conditioned on the measurement is
\begin{align}
    \Pr(\omega|d) & = \frac{\Pr(d|\omega)\Pr(\omega)}{\Pr(d)},
\end{align}
where $\Pr(d)$ is a normalization factor and $\Pr(d|\omega)$ is the likelihood of the experiment reproducing the observation $d$ given that the phase $\omega t$ was the true eigenphase of $U(t)$.

The art of Bayesian phase estimation is then to choose experiments in such a way to minimize the resources needed to reduce the uncertainty in an estimate of $\omega$, given by the estimator $\hat \omega$~\cite{svore_faster_2013,wiebe_efficient_2015}.
Adaptive versions of Bayesian phase estimation are known to achieve Heisenberg limited scaling and come close to saturating lower bounds on the uncertainty in $\omega$ as a function of experimental time or number of measurements~\cite{berry_optimal_2001,wiebe_efficient_2015}.
A complication that arises in choosing a Bayesian method is that each such method requires different amounts of experimental time, total number of measurements, classical memory and processing time and robustness to experimental imperfections.
The latter two are especially important for present-day experiments where any form of Bayesian inference that requires classical processing time that is comparable to the coherence time (which is often on the order of microseconds).
To this end, finding simple phase estimation methods that are robust, efficient and can be executed within a timescale of hundreds of nanoseconds remains an important problem.

\begin{figure}
    \begin{equation*}
        \Qcircuit @C=1em @R=1em {
            \lstick{\ket{0}}      & \gate{H}  & \gate{R_z(-t \omega_\inv)} & \ctrl{1}    & \gate{H} & \meter & \cw \\
            \lstick{\ket{\omega}} & {/} \qw   & \qw                        & \gate{U(t)} & \qw      & \qw    & \qw
        }
    \end{equation*}
    \caption{
        \label{fig:phase-est-circuit}
        Iterative phase estimation circuit used to measure a single datum in the experimental data set used to estimate $\omega$.
    }
\end{figure}
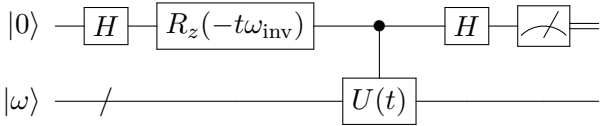

The likelihood function $\Pr(d|\omega)$ used in iterative phase estimation is given by the quantum circuit provided in~\autoref{fig:phase-est-circuit}.
Computing the probability of observing $d \in \{0,1\}$ as an outcome in the circuit shown in \autoref{fig:phase-est-circuit} gives us the \emph{likelihood function} for phase estimation,
\begin{align}
    \Pr(d | \omega; t, \omega_\inv) & = \cos^2(t [\omega - \omega_\inv] / 2+d\pi/2).\label{eq:likelihood}
\end{align}
Equipped with a likelihood function, we can thus reason about the posterior probability $\Pr(\omega | d)$ for a datum $d$ by using Bayesian inference.

In this setting we assume that we are able to apply the unitary $U(t)$ for any real-valued $t$.  This assumption is eminently reasonable in quantum simulation wherein, up to small errors on the order of the simulation error incurred in implementing $U(t)$, such unitaries can be constructed within sufficiently small error.  In settings where $U(t)$ can only can only be implemented directly for integer valued $t$, an evolution can be closely approximated using a quantum singular value transformation that is an $\epsilon$-approximation to $U^{t - \lfloor t \rfloor}$ using $O(\log(1/\epsilon))$ applications of $U(\lfloor t\rfloor )$~\cite{gilyen2019quantum} provided that $\|\log(U(t)) \|\le \pi/4$, which is typical for applications in quantum simulation.  These observations motivate our assumption that $U(t)$ is a continuous function of $t$.

A major challenge that we face when trying to numerically implement Bayesian phase estimation arises from the fact that we need to discretize the posterior distribution to ensure that an update to the distribution can be made in finite time.  Fortunately, unlike many other problems in Bayesian inference, the curse of dimensionality does not usually appear because the prior distribution $\Pr(\omega)$ maps $\mathbb{R}$ to $\mathbb{R}$.  Three natural approaches emerge when trying to model the posterior distribution are Sequential Monte-Carlo, Grids and Gaussian processes. 

Sequential Monte-Carlo is a general approach to Bayesian inference that approximates the posterior distribution as a sum of Dirac-delta functions: $\Pr(\omega) \approx \sum_j w_j \delta(\omega-\omega_j)$ for a set of phases $\{\omega_j:j=1,\ldots,N_{part}\}$.  As the posterior distribution is updated the $\omega_j$ are moved through a resampler to allow the distribution to continue to capture the low-order moments of the distribution even as the learning process excludes an exponentially growing fraction of the original prior distribution.  These Sequential-Monte-Carlo approaches been a workhorse for both phase estimation and also Hamiltonian learning: \cite{paesani_experimental_2017,wiebeQuantumBootstrappingCompressed2015}.  However, for multimodal distributions these methods can fail and since $N_{part} \in O(1/\epsilon^2)$ for most applications, this means that the classical memory requirements of storing the posterior distribution can be exponentially worse than one may expect from its contemporaries.  This makes such approaches less well suited for memory limited environments and motivated the development of rejection sampling~\cite{wiebe_efficient_2015} PE.

The posterior distribution is also discretized over grids in other works has also considered~\cite{svore_faster_2013,tipireddy2020bayesian}.  These techniques can be fast and accurate when used with adaptive grid refinement.  They fail however when the posterior distribution is not sufficiently smooth and suffer from the curse of dimensionality as all similar methods do.

The final approach that is broadly used, and mirrors the strategy taken here, exploits Gaussian processes to make the inference fast and highly memory efficient.  The central idea behind such methods is that they assume that the underlying prior distribution is a Gaussian and under the conditions that the likelihood function is sharply peaked the resulting posterior distribution will also be approximately Gaussian.  Since these updates map Gaussians to Gaussians using closed form expressions for the update they are highly efficient both in space and time and these methods form the most popular approaches to doing Bayesian inference in general.  However, in Heisenberg-limited phase estimation the underlying distribution is seldom Gaussian and so these methods have not seen as much use for generic phase estimation experiments (although similar ideas have been considered for non-Heisenberg limited phase estimation~\cite{lumino2018experimental}).


\section{Random Walk Phase Estimation}

\begin{algorithm}[H]
    \caption{\label{alg:rwpe-no-unwinding}
        Basic random walk phase estimation algorithm.
    }
    \begin{algorithmic}        
        \Function{RandomWalkPhaseEst}{$\mu_0$, $\sigma_0$}
            \Arguments
                \State $\mu_0$: initial mean
                \State $\sigma_0$: initial standard deviation
            \EndArguments
            \seccomment{Initialization}
            \State $\mu \gets \mu_0$
            \State $\sigma \gets \sigma_0$
            \seccomment{Main body}
            \For{$\iexp \in \{0, 1, \dots \nexp - 1\}$}
                \State $\omega_{\inv} \gets \mu - \pi \sigma / 2$
                \State $t \gets 1 / \sigma$
                \State Sample a datum $d$ from $\Pr(d = 0 | \omega; \omega_\inv, t) = \cos^2(t (\omega - \omega_\inv) / 2)$.
                \If{$d = 0$}
                    \State $\mu \gets \mu + \sigma / \sqrt{e}$
                \Else
                    \State $\mu \gets \mu - \sigma / \sqrt{e}$
                \EndIf
                \State $\sigma \gets \sigma \sqrt{(e - 1) / e}$
            \EndFor
            \seccomment{Final estimate}
            \State \Return $\hat{\omega} \gets \mu$
        \EndFunction
    \end{algorithmic}
\end{algorithm}

The central difference between our approach and most other Bayesian methods that have been proposed is that our method is entirely deterministic.
Specifically, the posterior mean (which we use as our estimate of the true eigenphase) shifts left or right upon collection of each datum by a fixed amount that depends only on the experimental outcome.
Thus the trajectory that our estimate of the eigenphase takes as the experiment proceeds follows a random walk with exponentially shrinking stepsize.
This simplicity allows us to not only store an approximation to the posterior distribution using shockingly little memory; it also only uses basic arithmetic that can be performed within the coherence times of modern quantum devices.

Rather than performing exact Bayesian inference, which is not efficient because the memory required grows exponentially with the number of bits of precision needed, we use a form of approximate Bayesian inference.
In particular, we assume that the prior distribution is Gaussian at each step:
\begin{align}
    \Pr(\omega) = \frac{\exp(-(\omega -\mu)^2/2\sigma^2)}{\sqrt{2\pi}\sigma}.
\end{align}
Such priors can be efficiently represented because they are only a function of $\mu$ and $\sigma$.
Unfortunately, the conjugate priors for~\autoref{eq:likelihood} are not Gaussian which means that the posterior distribution is not Gaussian:
\begin{widetext}
\begin{align}
    \Pr(\omega | d; t, \omega_\inv) & =
        \frac{
            \cos^2(t [\omega - \omega_\inv] / 2+d\pi/2)
            {\exp(-(\omega -\mu)^2/2\sigma^2)}
        }{\int_{-\infty}^\infty \cos^2(t [\omega - \omega_\inv] / 2+d\pi/2){\exp(-(\omega -\mu)^2/2\sigma^2)}\mathrm{d}\omega
            }.
\end{align}
\end{widetext}
However, for most experiments that we consider the posterior distribution will be unimodal and thus the majority of the probability mass will be given by the first two moments of the distribution.
This justifies the following approximation,
\begin{align}
    \Pr(\omega|d;t,\omega_\inv) \approx \frac{\exp(-(\omega -\mu')^2/2{\sigma'}^2)}{\sqrt{2\pi}\sigma'},
\end{align}
where $\mu'$ and $\sigma'$ are chosen to match the posterior mean and posterior standard deviation.
Specifically,
$$
    \mu' \defeq \expect[\omega] = \int_{-\infty}^\infty \omega P(d|\omega;t,\omega_\inv)\mathrm{d}\omega
$$
and similarly
$$
    {\sigma'}^2 \defeq \mathbb{V}[\omega] = \int_{-\infty}^\infty \omega^2 P(d|\omega;t,\omega_\inv)\mathrm{d}\omega - {\mu'}^2.
$$

Here, we will take as an approximation that the prior distribution $\Pr(\omega)$ is a Gaussian distribution at each step, and thus the prior can be characterized by its mean $\mu_0 \defeq \expect[\omega]$ and variance $\sigma_0^2 \defeq \mathbb{V}[\omega]$.
The Bayes update then consists of finding the updated mean $\mu = \expect[\omega | d] = \int \omega \Pr(\omega | d)\,\dd\omega$ and updated variance $\sigma^2 = \mathbb{V}[\omega | d] = \int \omega \Pr(\omega | d)\,\dd\omega - \mu^2$, which we can then use as the prior distribution at the next step.
For the phase estimation likelihood and under the assumption of prior Gaussianity, we can do this explicitly.
In calculating the update, we assume without loss of generality that under the prior, $\mu_0 = 0$  and $\sigma_0^2 = 1$; the general case is obtained from this case by the standard location--scale transformation.
We will express our update rule explicitly without the use of rescaling at the end of our derivation.
Without further ado, then, the rescaled update rule is given by
\begin{subequations}
    \label{eq:meanvar-updates}
    \begin{align}
        \mu' \gets & \frac{
            (t \sin[t \omega_\inv])
        }{
            s \e^{t^2/2} + \cos[t \omega_\inv])
        }
        \quad \text{ and }
        \\
        {\sigma'}^2 \gets & 1 - s t^2 \left[
            \frac{
                \e^{t^2 / 2} \cos(t \omega_\inv) + s
            }{
                \left(\e^{t^2/2} + s \cos(t \omega_\inv)\right)^2
            }
        \right],
    \end{align}
\end{subequations}
where $s = (-1)^d$.

Since the posterior variance describes the error that we should expect we will incur in estimating $\omega$ with the currently available observations, we can then choose $t$ and $\omega_\inv$ to minimize the posterior variance and hence minimize the errors that we will incur.
The posterior variance is minimized for both $d = 0, 1$ at $t = \sigma^{-1}$ and $\omega_\inv = \mu$, such that we can specialize \autoref{eq:meanvar-updates} at these values to obtain a simpler update rule,
\begin{subequations}
    \label{eq:opt-meanvar-updates}
    \begin{align}
        \mu' \gets & \mu + (-1)^d \frac{\sigma}{\sqrt{e}} \\
        \sigma' \gets & \sigma \sqrt{\frac{\e - 1}{\e}}
    \end{align}
\end{subequations}
In this way, we can \emph{derive} the particle guess heuristic (PGH) of \citet{wiebe_hamiltonian_2014}~and \citet{wiebeQuantumBootstrappingCompressed2015}~for the special case of phase estimation with a Gaussian conjugate prior. In particular, the PGH selects each new experiment by drawing two samples $\omega$ and $\omega'$ from the current posterior, then choosing $\omega_\inv = \omega$ and $t = 1 / \|\omega - \omega'\|$, such that minimizing \autoref{eq:meanvar-updates} agrees with the PGH in expectation over $\omega$ and $\omega'$.

Critically, the variance update in \autoref{eq:opt-meanvar-updates} does not depend on which datum we observe; the reduction in variance at each step is wholly deterministic.
Similarly, the mean shifts by the same amount at each step, such that the only dependence on the data of the Gaussian approximation to the posterior distribution is through which \emph{direction} the mean shifts at each observation.
We can thus think of the Gaussian approximation as yielding a random walk that is damped towards the true value of $\omega$.  This leads to two observations:
\begin{enumerate}
    \item The update rule for the position of the posterior Gaussian distribution is translationally invariant and scale invariant.
    \item The update rule for the posterior variance is scale invariant.
\end{enumerate}
Following this identification, we obtain \autoref{alg:rwpe-no-unwinding}, which exploits these invariances in the approximate posterior density to render the inference step computationally trivial and therefore suitable for rapid execution in both ASICs and FPGAs.

\begin{figure*}
    \begin{center}
        \includegraphics[width=0.9\textwidth]{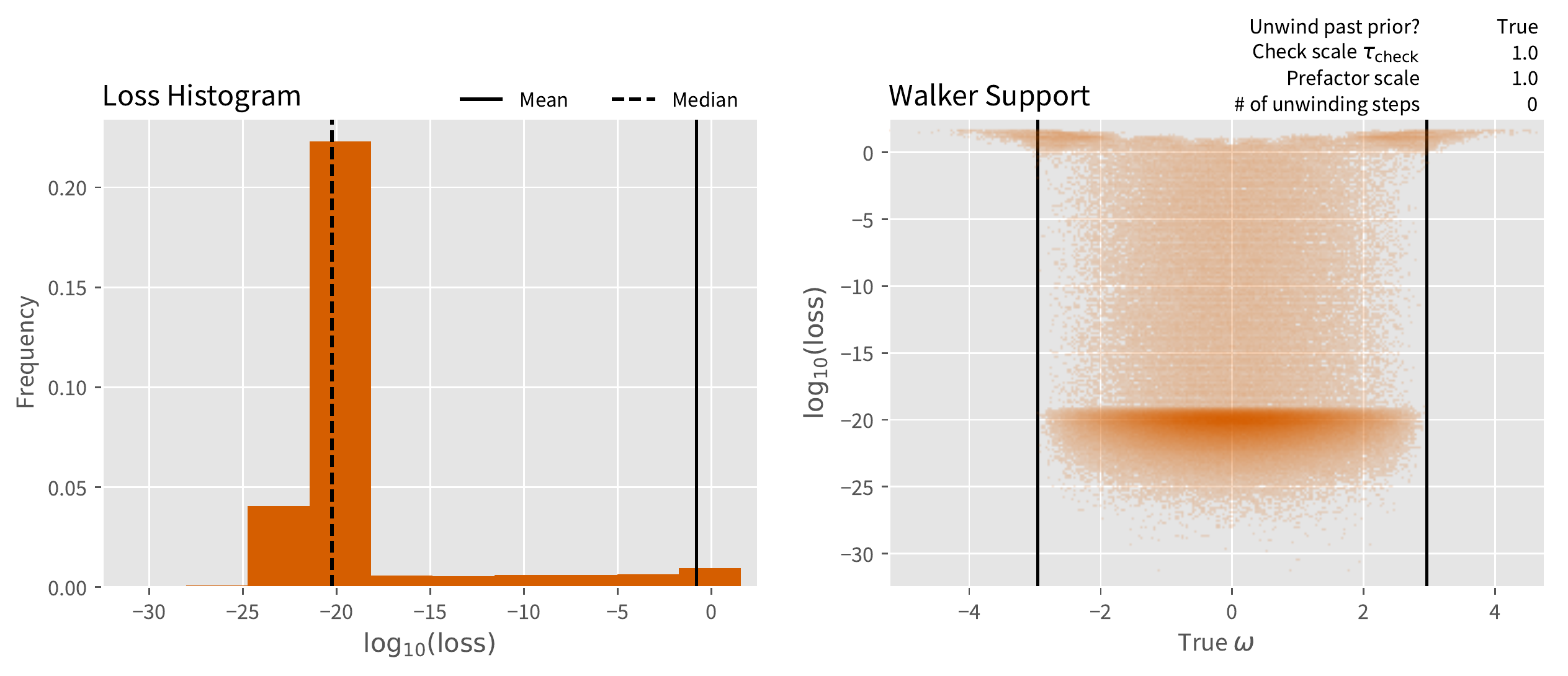}

        \includegraphics[width=0.9\textwidth]{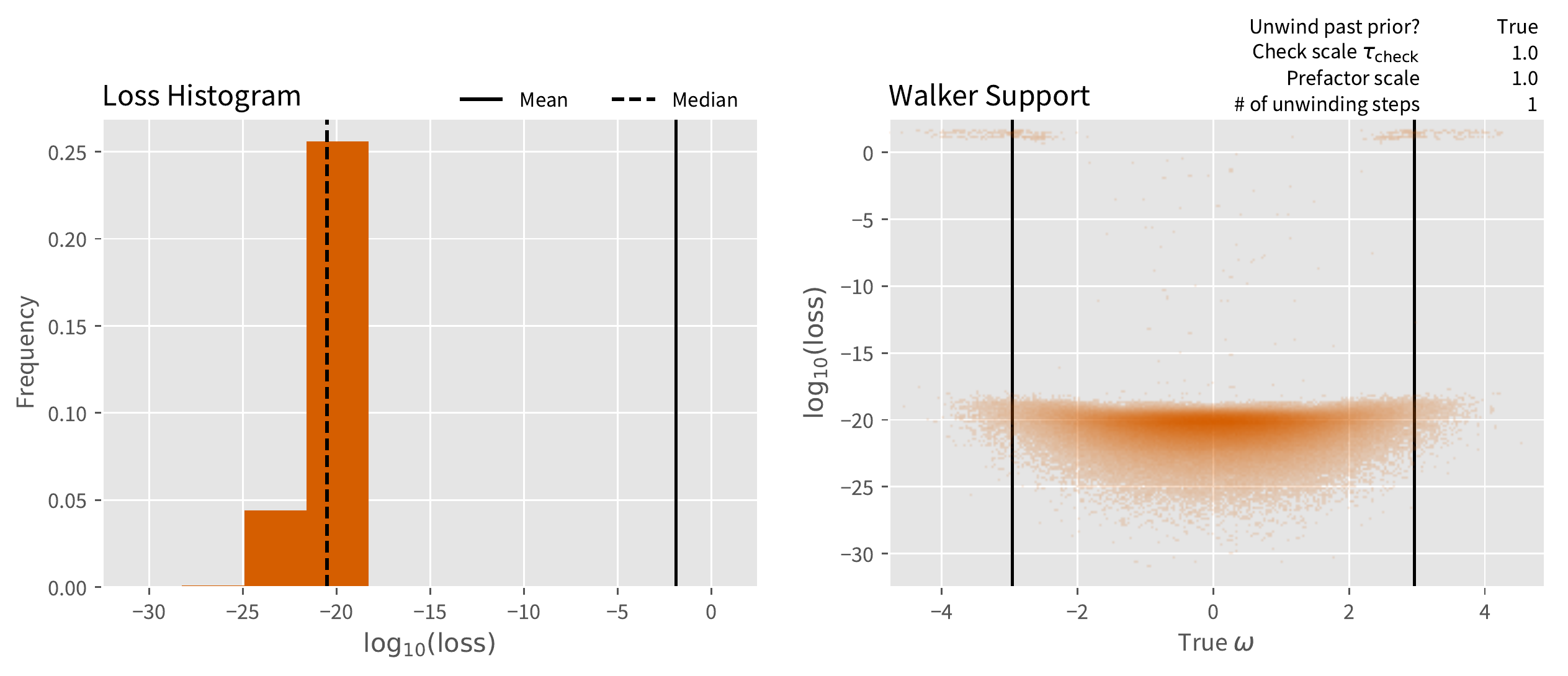}

        \includegraphics[width=0.9\textwidth]{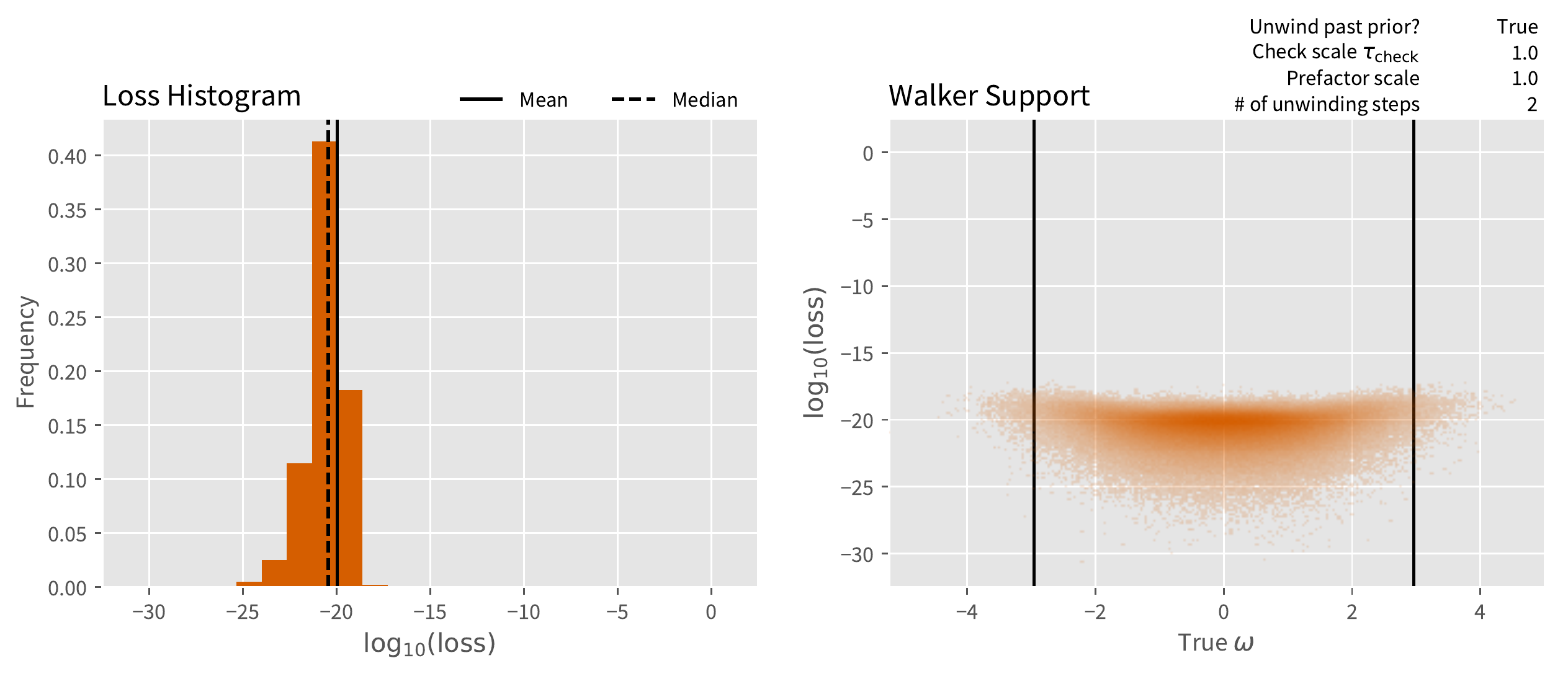}
    \end{center}
    \caption{
        \label{fig:alg-loss-histogram}
        \textbf{(Left column)}
            Histogram over log-losses for many different trials, compared to the mean loss (Bayes risk)
            and the median loss.
        \textbf{(Right column)}
            Log-loss versus the true value of $\omega$, compared with the finite range (thick lines)
            in which each walker can explore.
        \textbf{(Top row)}
            Trials are run using the basic approach of \autoref{alg:rwpe-no-unwinding}.
        \textbf{(Middle row)}
            Trials are run using one unwinding step, as in \autoref{alg:rwpe-w-unwinding}.
        \textbf{(Bottom row)}
            Trials are run using two unwinding steps, as in \autoref{alg:rwpe-w-unwinding}.
    }
\end{figure*}

\begin{figure*}
    \begin{center}
        \includegraphics[width=0.475\textwidth]{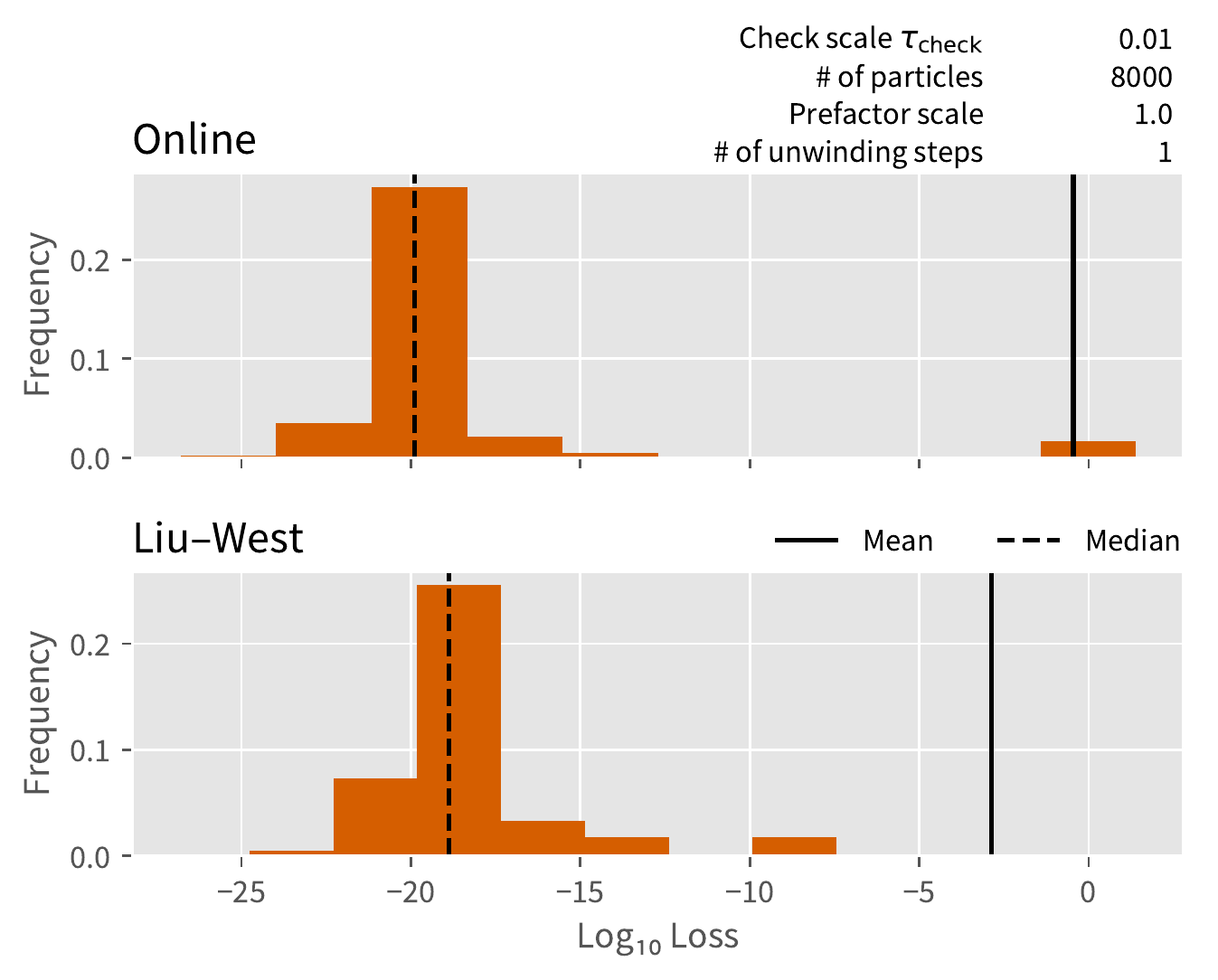}
        \includegraphics[width=0.475\textwidth]{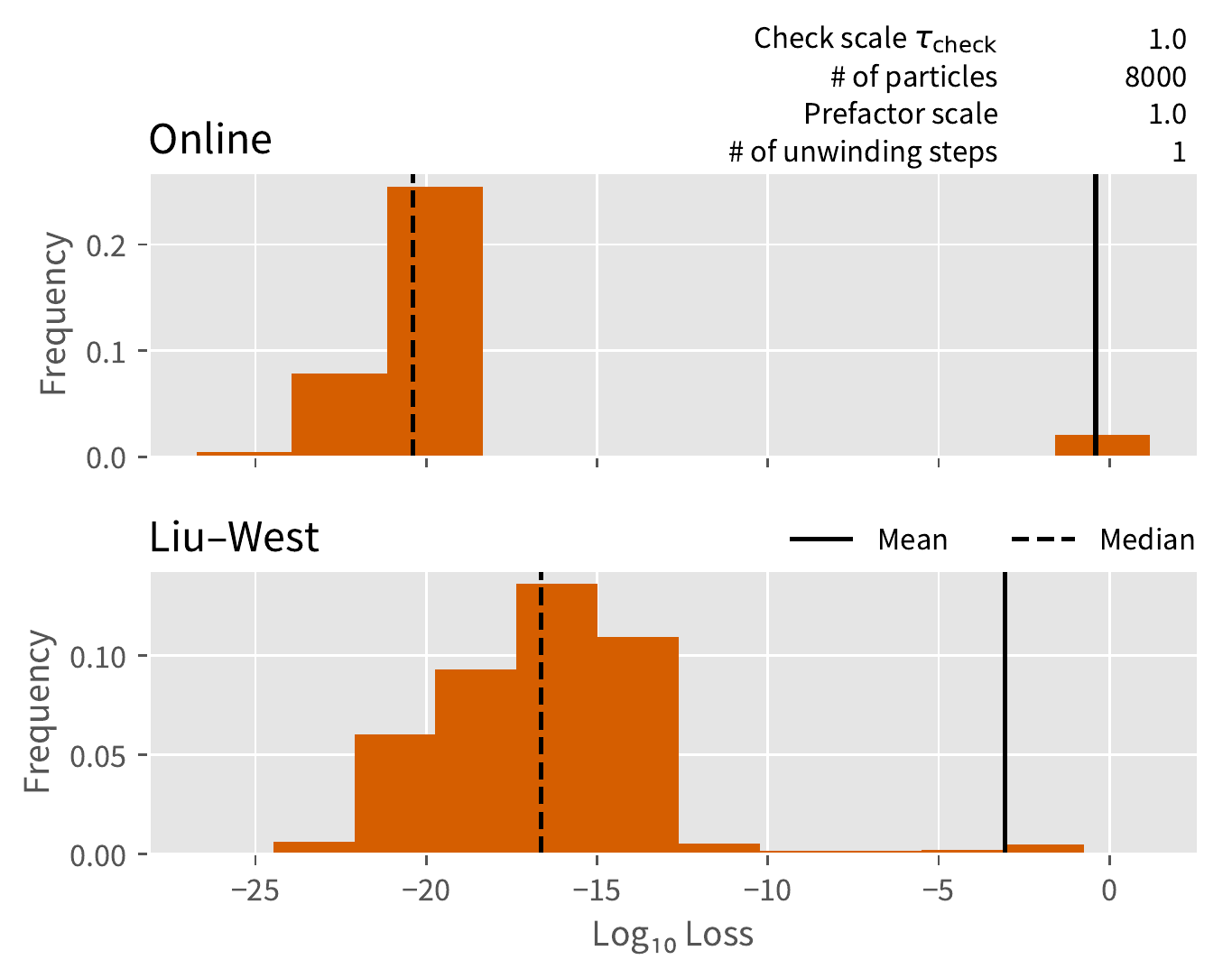}
    \end{center}
    \caption{
        \label{fig:online-v-lw-hist}
        \textbf{(Left column)}
            $\tch = 10^{-2}$
        \textbf{(Right column)}
            $\tch = 10^0$
        \textbf{(Top row)}
            Histogram over log-losses for many different trials, using \autoref{alg:rwpe-w-unwinding}
            with $n_\unwind = 1$.
        \textbf{(Bottom row)}
            Histogram over log-losses for the same trials as in the top row, but using particle
            filtering with Liu--West resampling to incorporate consistency check and unwound data.
    }
\end{figure*}


\autoref{alg:rwpe-no-unwinding} can fail to provide an inaccurate estimate of $\hat{\omega}$ in two different ways.
First, that the mean can only shift by fixed amount at each step implies that the walk on $\omega$ can not go further than
\begin{align}\label{eq:RWPE_range}
    \frac{1}{\sqrt{e}} \sum_{k=0}^{\infty} \left(\frac{e - 1}{e} \right)^{k / 2} = \frac{1}{\sqrt{e} - \sqrt{e - 1}} \approx 2.95
\end{align}
away from its initial position.
Assuming $\mu_0 = 0$ and $\sigma_0 = 1$ correctly describe the true probability distribution from which $\omega$ are chosen, this corresponds to a failure probability of 0.3\%.
If we take a prior that is too ``tight'' by a factor of about 20\% (corresponding to a single step of the algorithm), then the failure probability due to finite support of the random walk can be as large as 1.8\%.
Thankfully, this is an easy failure modality to address, in that we can use an initial prior distribution with an artificially wider variance than our actual prior.
We confirm this intuition in~\autoref{fig:alg-loss-histogram} wherein we see that these failures tend to lead to rare events where improbable sequences of measurements that causes the mean error to be much larger than the median error due to rare but significant deviations from the true value.

The other failure modality is more severe, in that the Gaussian approximation itself can fail.
By contrast to rejection filtering and other such approaches, in random walk phase estimation, we do not choose the variance based on the data itself, but based on an offline analytic computation.
Thus, we must have a way to correct for approximation failures.
We propose to do this by \emph{unwinding} data records, returning to an earlier state of the random walk algorithm.
In particular, since the data record uniquely describes the entire path of a random walker, we can step backwards along the history of a random walk.

To determine when we should unwind in this fashion, we will perform additional measurements whose outcomes we can accurately predict if the approximation is still accurate.
Choosing $\omega_\inv = \mu$ and $t = \tch / \sigma$ for a scale parameter $\tch > 0$, we obtain that the probability of a 0 outcome is
\begin{subequations}
    \begin{align}
        \Pr(0; \mu, \tch / \sigma) & = \int \Pr(0 | \omega; \mu, \tch / \sigma) \Pr(\omega) \dd\omega \\
                                   & = \frac12 \left(
                                       1 + e^{-\tch^2 / 2}
                                   \right),
    \end{align}
\end{subequations}
where $\Pr(\omega)$ is assumed to be the normal distribution with mean $\mu$ and variance $\sigma^2$ as per the Gaussianity assumption.
For $\tch = 1$, this corresponds to a false negative rate (that is, inferring the approximation has failed when it has not) of approximately $19.6\%$, while $\tch = 10^{-2}$ gives a false negative rate of $0.002\%$.
The key idea behind this is to use a low-cost consistency check to determine whether the Gaussian assumptions for the posterior distribution still hold and if they fail then we revert the random walk until the consistency check passes. 
We list the unwinding together with the original random walk in \autoref{alg:rwpe-w-unwinding}.

Notably, both the consistency check data and the unwound data are still useful data.
In situations where the data can be efficiently recorded for later use, the entire data record including all unwound data and consistency checks may be used in postprocessing to verify that the estimate derived from the random walk is feasible.
In the numerical results presented in \autoref{sec:numerics}, we use the example of particle filtering to perform postprocessing \cite{doucet_tutorial_2011}, using the QInfer software library \cite{granade_qinfer_2017}.

\section{Numerical Results}
\label{sec:numerics}
The data given in~\autoref{fig:alg-loss-histogram} provides an indication of the performance of RWPE for typical phase estimation experiments.  We find from the data provided that RWPE works  extremely well over the trials considered.  Specifically, we take our initial value of $\omega$ to be Gaussian distributed according to the initial prior of the distribution.  Note that the data as seen from the true $\omega$ distribution is not distributed via a circular distribution but rather a linear Gaussian distribution.  While this creates some bias in the output angles in principle, the distribution is broad enough that it approximates a uniform distribution here.

More concretely, we see that for experiments consisting of at most $100$ accepted steps for the random walk (and a maximum of $100~000$ total steps) that the median loss for RWPE is low.  Specifically, it is on the order of $10^{-20}$ which corresponds to $10$ digits of accuracy.  Also, as a point of comparison the Heisenberg limit for the phase observed is $3.5\times 10^{-11}$~\cite{berry_optimal_2001}.  We find from RWPE that the median performance is close to the Heisenberg limit, specifically the median is also on the order of $3\times 10^{-11}$; however, a direct comparison to the Heisenberg limit may not be fully appropriate as the definition is usually formally given in terms of the Holevo variance rather than the median deviation.  These differences are apparent in the mean, where we find that the distribution of errors in~\autoref{fig:alg-loss-histogram} has losses that are on the order of $1$ with non-negligible frequency for the case where no unwinding steps are taken.  The impact of this is severely limited by taking one unwinding step; however, we see that for the examples considered a single case of failure (which does not appear in the histogram at the scale presented) causes the mean error to be substantial.  Despite the median being small, the large mean errors in both cases suggest that we formally are not Hesienberg limited.  In contrast, two unwinding steps were found to completely eliminate these issues and the mean loss was observed to shrink to $10^{-20}$, which yields a variance that is within less than an order of magnitude of the Heisenberg limit for these experiments.

The cause of many of these failures can be seen in the support of the errors over the true value of $\omega$.  If no unwinding is performed then the mean-square error is on the order unity for true frequencies that are close to $\pi$ or $-\pi$.  This is because the update rules for the random walk can only explore $[-2.95,2.95]$ according to~\eqref{eq:RWPE_range}.  The ability to unwind our variance past the initial prior allows the unwinding algorithm to deal with this.

We further see from the histogram of the quadratic loss for the reported eigenphase versus the actual eigenphase in~\autoref{fig:online-v-lw-hist} that the median performance of RWPE outperforms the results seen by Liu-West when applied to the same dataset.  We perform this part in order to ensure that even the data used in the consistency check is provided to both methods.  For the case where the check timescale is small, $\tau_{\rm check}=0.01$, we see that the median quadratic loss is nearly two orders of magnitude smaller than Liu-West using $8000$ particles.  For the case of the larger time used for the consistency check, $\tau_{\rm check} =1$, we find that there is a separation between the two on the order of nearly $5$ orders of magnitude.  The increased separation between the performance is likely because of the larger timesteps used in the consistency check step leading to multi-modal distributions in the particle filter which cannot be well approximated using the Liu-West resampler.  

In both cases the mean errors are observed to be large.  This is expected as when a failure occurs the error tends to be on the order of unity, which means that when the median is on the order of $10^{-10}$ that rare errors can dominate the mean, however they do not dominate the median.  These problems can be mitigated by increasing the number of rewinding steps or increasing the number of particles used in the particle filter approximations.

Here $8000$ particles are used in the particle filters because previous work~\cite{wiebe_efficient_2015} found that this tended to yield stable results while also allowing updates to be performed on a scale of tens of milliseconds on a single-core CPU.  In contrast, RWPE requires time on the order of a microsecond to perform an update on a single core CPU.  The difference between the two processing times means that RWPE is vastly better suited for online setting in superconducting hardware where the coherence times are often on the order of  $100\mu$s.

\begin{figure*}[t!]
    \begin{center}
        \includegraphics[width=0.475\textwidth]{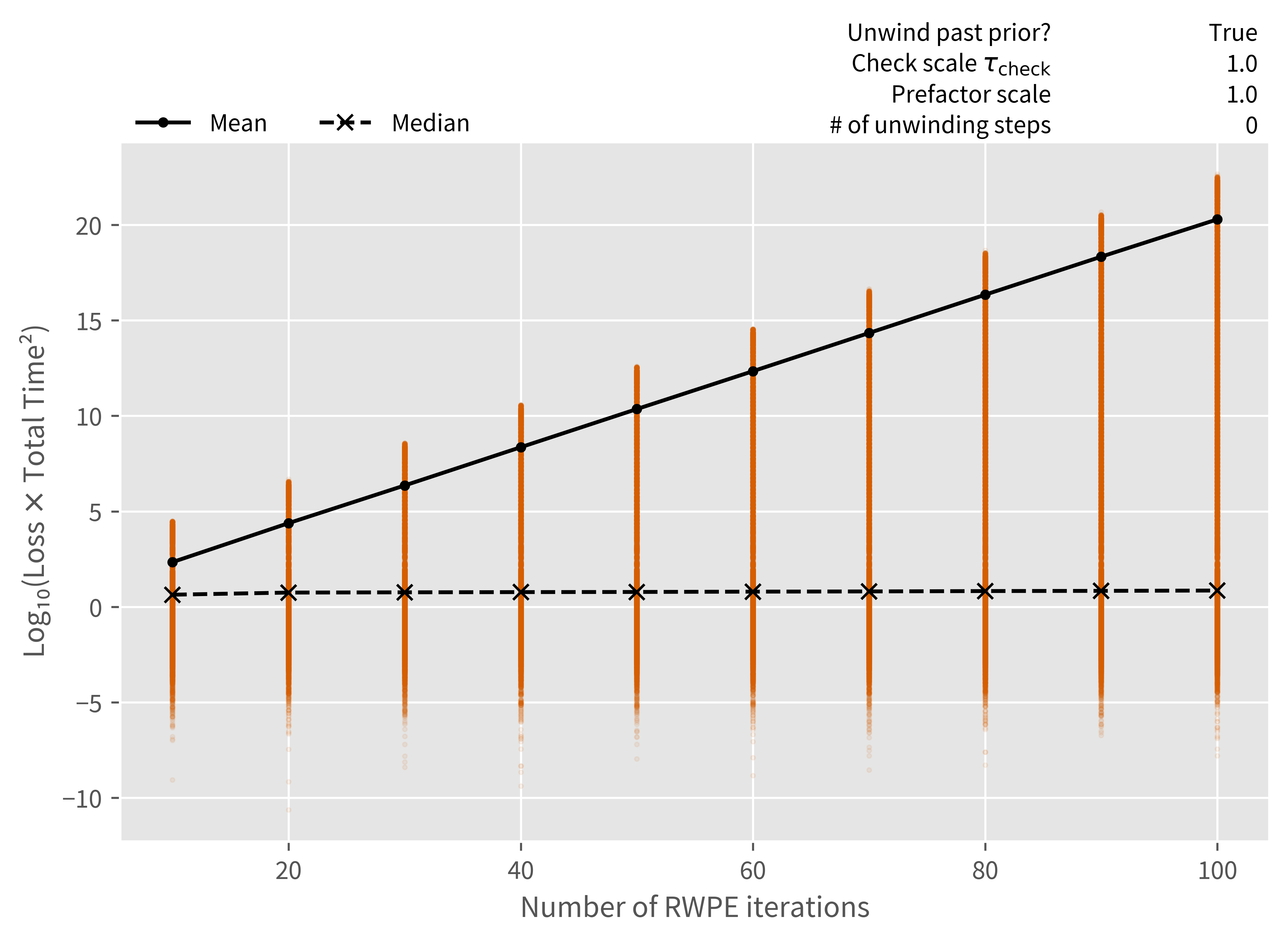}
        \includegraphics[width=0.475\textwidth]{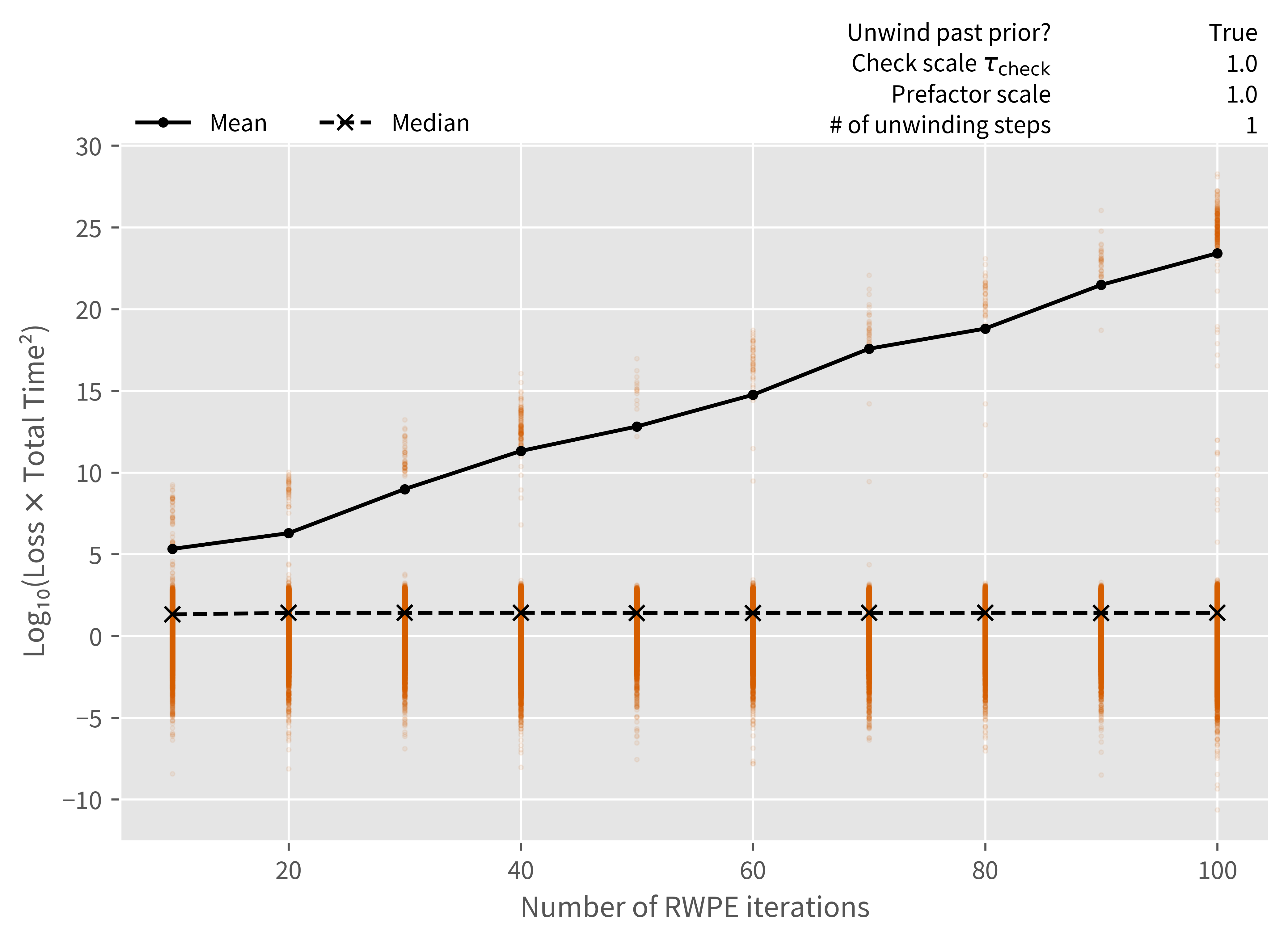}

        \includegraphics[width=0.475\textwidth]{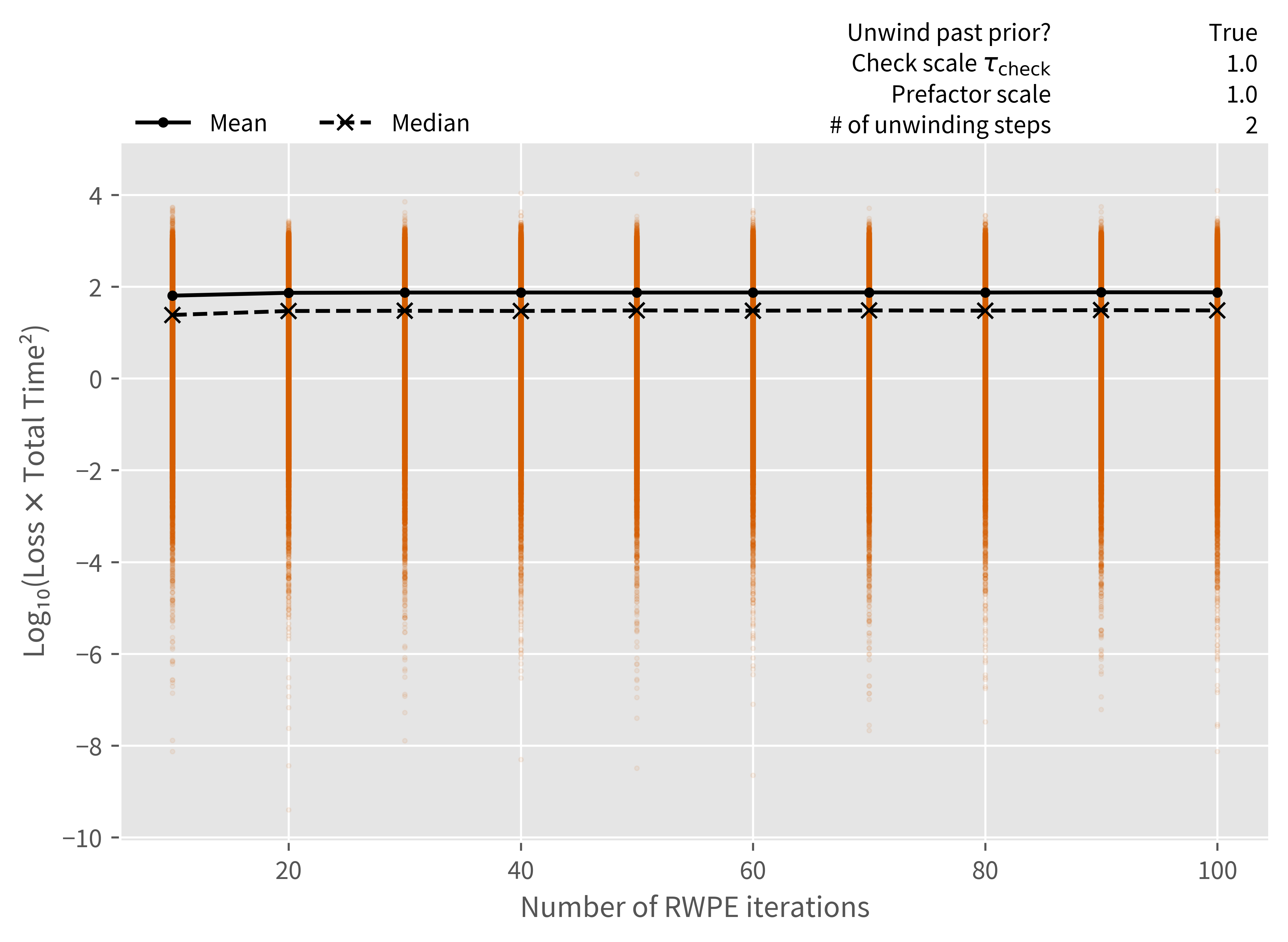}
        \includegraphics[width=0.475\textwidth]{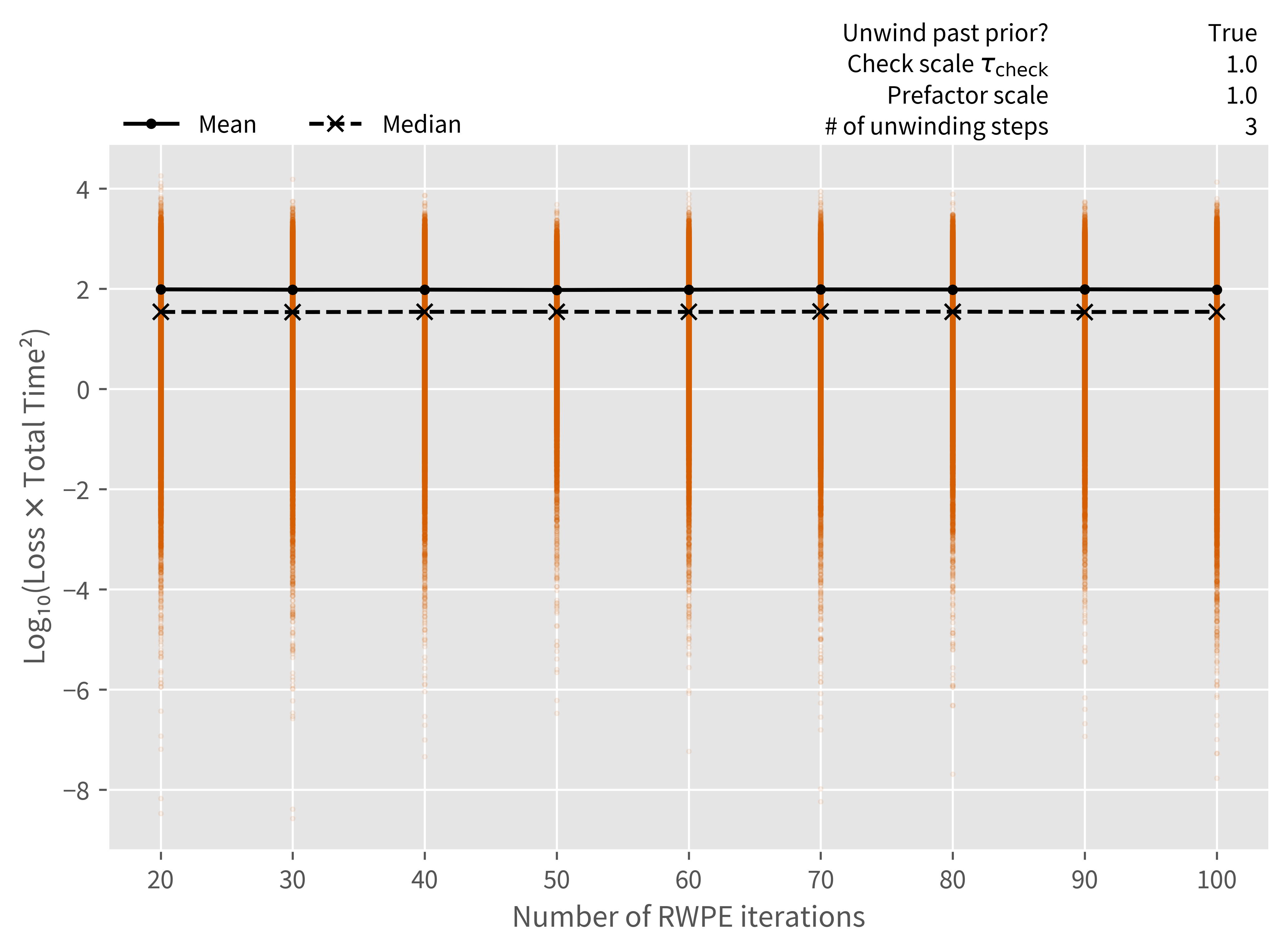}
    \end{center}
    \caption{
        \label{fig:heisenberg-vs-n-exp}
        The data presented here provide a demonstration of Heisenberg limited scaling, which corresponds to a quadratic loss that is inverse to the square of the time for up to $100$ accepted steps for RWPE. Here the Heisenberg limit corresponds to a line of zero slope for the mean square error and is observed for the data with $2$ and $3$ unwinding steps.
    }
\end{figure*}

The final question that we wish to probe numerically is whether the algorithm is capable of achieving Heisenberg limited scaling in practice.  We defer discussion about the optimal constants for the scaling to the following section.  Specifically, for Heisenbert limited scaling we need to have that the Holevo variance scales like the inverse of the square of the total simulation time.  The Holevo variance corresponds to the ordinary variance for narrow distributions that have negligible overlap with the branch cut chosen for the eigenphases returned by phase estimation.

We examine this scaling in~\autoref{fig:heisenberg-vs-n-exp}.  We find that for zero and one unwinding step, that while the median error is small, the mean error does not shrink as the number of iterations  grows.  In contrast, we see clear evidence (a line of zero slope) for Heisenberg limited scaling if we use two or three unwinding steps.  This suggests that the processing technique that RWPE is capable of achieving Heisenberg limited scaling even in an online setting.  




\section{Decision Theoretic Bounds}

\newrm{BME}

In this work, we have expressed our algorithm as approximating the Bayesian mean estimator (BME) $\hat{\omega}_\BME(D) \defeq \expect[\omega | D]$.
This choice is motivated by a decision theoretic argument, in which we penalize an estimator by a particular amount $L(\hat{\omega}(D), \omega)$ for returning a particular estimate $\hat{\omega}(D)$ when the true value of the parameter of interest is $\omega$.
A function $L$ which we use to assign such penalties is called a \emph{loss function}; as described in the main body, the quadratic loss $L_Q(\hat{\omega}(D) - \omega) \defeq (\hat{\omega}(D) - \omega)^2$ is a particularly useful and broadly applicable loss function.
We will drop the subscript $Q$ when it is clear from context that we are referring to this choice.

In any case, once we have adopted such a loss function, we can then reason about the average loss incurred by a particular estimator.
In particular, the \emph{Bayes risk} $r$ is defined as the expected loss over all hypotheses $\omega \sim \pi$ drawn from a prior $\pi$ and over all data sets $D$.
That is,
\begin{align}
    r(\hat{\omega}) \defeq \expect_{\omega \sim \pi, D}[L(\omega(D), \omega)].
\end{align}
The celebrated result of \citet{banerjeeOptimalityConditionalExpectation2005}~then provides that the BME minimizes the Bayes risk for any loss function of the form $L(\hat{\omega}(D), \omega) = F(\hat{\omega}(D)) - F(\omega) - (\frac{\dd}{\dd\omega} F(\omega)) (\hat{\omega} - \omega)$ for a convex function $F$.
This holds for the quadratic loss, as well as a number of other important examples such as the Kullback--Leibler divergence.

Having thus argued that our approach is approximately optimal for the Bayes risk, we are then interested in how \emph{well} our algorithm approximately achieves the Bayes risk.
To address this, we note that the van Trees inequality \cite{gillApplicationsVanTrees1995}~lower bounds the Bayes risk achievable by any estimator $\hat{\omega}$ as
\begin{align}
    r(\hat{\omega}) \ge \frac{1}{\expect_{\omega \sim \pi}[I(\omega)] + I_0},
\end{align}
where $I(\omega) \defeq \expect_{D}[\left(\frac{\dd}{\dd \omega} \log \Pr(D | \omega) \right)^2]$ is the usual Fisher information, and where $I_0 \defeq \expect_{\omega \sim \pi}[\left(\frac{\dd}{\dd\omega} \log \pi(\omega)\right)^2]$ is a correction for the information provided by the prior distribution $\pi$.
Thus, the van Trees inequality can be thought of as a correction to the traditional Cram\'er--Rao bound for the effects of prior information~\cite{cover1991information}.
Whereas the Cram\'er--Rao bound only holds for unbiased estimators, the BME is asymptotically unbiased but may in general be biased for a finite-sized data set.

In the case that $\pi$ is a normal distribution $\mathcal{N}(0, 1)$, however, the correction $I_0$ for the bias introduced by the prior distribution is identically zero, such that the van Trees and Cram\'er--Rao bounds coincide.
More generally, the set of true values $\omega$ for which the Cram\'er--Rao bound can be violated by a biased estimator approaches Lebesgue measure zero asymptotically with the number of observations $\nexp$ \cite{Opper:1999:BAO:304710.304756}, such that the Cram\'er--Rao bound is useful for characterizing the performance of biased estimators such as the BME.

Applying this bound to \autoref{alg:rwpe-w-unwinding}, we note that the experiments chosen by the algorithm are \emph{deterministic} unless the unwinding step proceeds backwards from the initial prior.
Thus, the Cram\'er--Rao and hence the van Trees inequalities can be calculated explicitly for the particular experimental choices made by \autoref{alg:rwpe-w-unwinding}, yielding that
\begin{align}
    \label{eq:van-trees-final}
    r(\hat{\omega}) \ge \frac{1}{\sum_{\iexp = 0}^{\nexp - 1} \left(\frac{\e}{\e - 1}\right)^{\iexp}}
                    = \left(
                        \left[\e - 1\right] \left[\frac{\e}{\e - 1}\right]^{\nexp}
                    \right)^{-1},
\end{align}
where we have used that $I(\omega) = \sum_i t_i^2$ \cite{ferrieHowBestSample2013}.
For $\nexp = 100$, $r(\hat{\omega}) \gtrapprox 7 \times 10^{-21}$.  This corresponds to the mean-square error observed in \autoref{fig:alg-loss-histogram} for the case where $2$ unwinding steps is used and thus RWPE performs a nearly optimal analysis of the data using on the order of a millisecond of total processing time.

Critically, the van Trees inequality holds in expectation, such that the loss in any \emph{particular} experiment can be significantly better than \autoref{eq:van-trees-final}.
Moreover, the consistency checks utilized by \autoref{alg:rwpe-w-unwinding} provide additional data not included in the bound given by \autoref{eq:van-trees-final}, such that our algorithm violate the van Trees inequality evaluated only at the accepted measurement results.
In that sense, our algorithm can be thought of as a heuristic for importance sampling from amongst the data, such that the usual discussions of optimality and postselection apply~\cite{ferrieHowBestSample2013,combes2014quantum}.




\section{Conclusion}
To conclude, our work has shown a new approach to adaptive Bayesian phase estimation that uses Gaussian processes coupled with an unwinding rule to improve the stability of the learning process.  Our method requires a constant amount of memory to perform the analysis and is further Heisenberg limited.  Further, the fact that a constant application needs to be performed at every step in the process means that it is trivial to implement the updates to the prior distribution here using an FPGA without resorting to a lookup table.  This makes our approach arguably the best known Heisenberg-limited approach for quantum phase estimation on near-term experiments wherein latency may be a concern and limitations due to heating power proves a bottleneck.  Specifically, we find that our approach to phase estimation requires $O(\log(1/\epsilon))$ bits of memory to perform phase estimation within error epsilon and requires $O(\log(1/\epsilon))$ gate operations to perform the phase estimation assuming a constant number of unwinding steps are needed if single-precision data types are used.  In contrast, Bayesian phase estimation using SMC requires $O(\log(1/\epsilon)/\epsilon^2)$ arithmetic operations and cosine evaluations to perform phase estimation~\cite{granade2012robust}.  This gives our approach a substantial advantage in classical time and space complexity relative to its classical brethren.

Important applications of these ideas include not only quantum phase estimation for algorithmic applications but also metrology applications and further these ideas are sufficiently cost effective that in some platforms with slow quantum operations, they could be used to assist in adaptive measurement of qubits or other quantum systems.  Further, while this work focuses on applications within the quantum domain, similar considerations also apply outside of this setting and the ability to do adaptive inference in memory limited environment may have further applications in broader settings such as with autonomous drones or in control engineering.  Regardless, the ideas behind this work show that adaptive Bayesian inference need not be expensive and through the use of clever heuristics can be brought within the reach of even the most modest control hardware.
\nocite{apsrev41Control}
\bibliographystyle{apsrev4-1}
\bibliography{apsrev-control,random-walk-phase-est}

\begin{thebibliography}{31}%
\makeatletter
\providecommand \@ifxundefined [1]{%
 \@ifx{#1\undefined}
}%
\providecommand \@ifnum [1]{%
 \ifnum #1\expandafter \@firstoftwo
 \else \expandafter \@secondoftwo
 \fi
}%
\providecommand \@ifx [1]{%
 \ifx #1\expandafter \@firstoftwo
 \else \expandafter \@secondoftwo
 \fi
}%
\providecommand \natexlab [1]{#1}%
\providecommand \enquote  [1]{``#1''}%
\providecommand \bibnamefont  [1]{#1}%
\providecommand \bibfnamefont [1]{#1}%
\providecommand \citenamefont [1]{#1}%
\providecommand \href@noop [0]{\@secondoftwo}%
\providecommand \href [0]{\begingroup \@sanitize@url \@href}%
\providecommand \@href[1]{\@@startlink{#1}\@@href}%
\providecommand \@@href[1]{\endgroup#1\@@endlink}%
\providecommand \@sanitize@url [0]{\catcode `\\12\catcode `\$12\catcode
  `\&12\catcode `\#12\catcode `\^12\catcode `\_12\catcode `\%12\relax}%
\providecommand \@@startlink[1]{}%
\providecommand \@@endlink[0]{}%
\providecommand \url  [0]{\begingroup\@sanitize@url \@url }%
\providecommand \@url [1]{\endgroup\@href {#1}{\urlprefix }}%
\providecommand \urlprefix  [0]{URL }%
\providecommand \Eprint [0]{\href }%
\providecommand \doibase [0]{http://dx.doi.org/}%
\providecommand \selectlanguage [0]{\@gobble}%
\providecommand \bibinfo  [0]{\@secondoftwo}%
\providecommand \bibfield  [0]{\@secondoftwo}%
\providecommand \translation [1]{[#1]}%
\providecommand \BibitemOpen [0]{}%
\providecommand \bibitemStop [0]{}%
\providecommand \bibitemNoStop [0]{.\EOS\space}%
\providecommand \EOS [0]{\spacefactor3000\relax}%
\providecommand \BibitemShut  [1]{\csname bibitem#1\endcsname}%
\let\auto@bib@innerbib\@empty
\bibitem [{\citenamefont {Shor}(1994)}]{shor1994algorithms}%
  \BibitemOpen
  \bibfield  {author} {\bibinfo {author} {\bibfnamefont {P.~W.}\ \bibnamefont
  {Shor}},\ }\bibfield  {title} {\enquote {\bibinfo {title} {Algorithms for
  quantum computation: discrete logarithms and factoring},}\ }in\ \href@noop {}
  {\emph {\bibinfo {booktitle} {Proceedings 35th annual symposium on
  foundations of computer science}}}\ (\bibinfo {organization} {Ieee},\
  \bibinfo {year} {1994})\ pp.\ \bibinfo {pages} {124--134}\BibitemShut
  {NoStop}%
\bibitem [{\citenamefont {Harrow}\ \emph {et~al.}(2009)\citenamefont {Harrow},
  \citenamefont {Hassidim},\ and\ \citenamefont {Lloyd}}]{harrow2009quantum}%
  \BibitemOpen
  \bibfield  {author} {\bibinfo {author} {\bibfnamefont {A.~W.}\ \bibnamefont
  {Harrow}}, \bibinfo {author} {\bibfnamefont {A.}~\bibnamefont {Hassidim}}, \
  and\ \bibinfo {author} {\bibfnamefont {S.}~\bibnamefont {Lloyd}},\ }\bibfield
   {title} {\enquote {\bibinfo {title} {Quantum algorithm for linear systems of
  equations},}\ }\href@noop {} {\bibfield  {journal} {\bibinfo  {journal}
  {Physical review letters}\ }\textbf {\bibinfo {volume} {103}},\ \bibinfo
  {pages} {150502} (\bibinfo {year} {2009})}\BibitemShut {NoStop}%
\bibitem [{\citenamefont {Reiher}\ \emph {et~al.}(2017)\citenamefont {Reiher},
  \citenamefont {Wiebe}, \citenamefont {Svore}, \citenamefont {Wecker},\ and\
  \citenamefont {Troyer}}]{reiher2017elucidating}%
  \BibitemOpen
  \bibfield  {author} {\bibinfo {author} {\bibfnamefont {M.}~\bibnamefont
  {Reiher}}, \bibinfo {author} {\bibfnamefont {N.}~\bibnamefont {Wiebe}},
  \bibinfo {author} {\bibfnamefont {K.~M.}\ \bibnamefont {Svore}}, \bibinfo
  {author} {\bibfnamefont {D.}~\bibnamefont {Wecker}}, \ and\ \bibinfo {author}
  {\bibfnamefont {M.}~\bibnamefont {Troyer}},\ }\bibfield  {title} {\enquote
  {\bibinfo {title} {Elucidating reaction mechanisms on quantum computers},}\
  }\href@noop {} {\bibfield  {journal} {\bibinfo  {journal} {Proceedings of the
  national academy of sciences}\ }\textbf {\bibinfo {volume} {114}},\ \bibinfo
  {pages} {7555} (\bibinfo {year} {2017})}\BibitemShut {NoStop}%
\bibitem [{\citenamefont {Abrams}\ and\ \citenamefont
  {Lloyd}(1997)}]{abrams1997simulation}%
  \BibitemOpen
  \bibfield  {author} {\bibinfo {author} {\bibfnamefont {D.~S.}\ \bibnamefont
  {Abrams}}\ and\ \bibinfo {author} {\bibfnamefont {S.}~\bibnamefont {Lloyd}},\
  }\bibfield  {title} {\enquote {\bibinfo {title} {Simulation of many-body
  fermi systems on a universal quantum computer},}\ }\href@noop {} {\bibfield
  {journal} {\bibinfo  {journal} {Physical Review Letters}\ }\textbf {\bibinfo
  {volume} {79}},\ \bibinfo {pages} {2586} (\bibinfo {year}
  {1997})}\BibitemShut {NoStop}%
\bibitem [{\citenamefont {von Burg}\ \emph {et~al.}(2021)\citenamefont {von
  Burg}, \citenamefont {Low}, \citenamefont {H{\"a}ner}, \citenamefont
  {Steiger}, \citenamefont {Reiher}, \citenamefont {Roetteler},\ and\
  \citenamefont {Troyer}}]{von2021quantum}%
  \BibitemOpen
  \bibfield  {author} {\bibinfo {author} {\bibfnamefont {V.}~\bibnamefont {von
  Burg}}, \bibinfo {author} {\bibfnamefont {G.~H.}\ \bibnamefont {Low}},
  \bibinfo {author} {\bibfnamefont {T.}~\bibnamefont {H{\"a}ner}}, \bibinfo
  {author} {\bibfnamefont {D.~S.}\ \bibnamefont {Steiger}}, \bibinfo {author}
  {\bibfnamefont {M.}~\bibnamefont {Reiher}}, \bibinfo {author} {\bibfnamefont
  {M.}~\bibnamefont {Roetteler}}, \ and\ \bibinfo {author} {\bibfnamefont
  {M.}~\bibnamefont {Troyer}},\ }\bibfield  {title} {\enquote {\bibinfo {title}
  {Quantum computing enhanced computational catalysis},}\ }\href@noop {}
  {\bibfield  {journal} {\bibinfo  {journal} {Physical Review Research}\
  }\textbf {\bibinfo {volume} {3}},\ \bibinfo {pages} {033055} (\bibinfo {year}
  {2021})}\BibitemShut {NoStop}%
\bibitem [{\citenamefont {Su}\ \emph {et~al.}(2021)\citenamefont {Su},
  \citenamefont {Berry}, \citenamefont {Wiebe}, \citenamefont {Rubin},\ and\
  \citenamefont {Babbush}}]{su2021fault}%
  \BibitemOpen
  \bibfield  {author} {\bibinfo {author} {\bibfnamefont {Y.}~\bibnamefont
  {Su}}, \bibinfo {author} {\bibfnamefont {D.~W.}\ \bibnamefont {Berry}},
  \bibinfo {author} {\bibfnamefont {N.}~\bibnamefont {Wiebe}}, \bibinfo
  {author} {\bibfnamefont {N.}~\bibnamefont {Rubin}}, \ and\ \bibinfo {author}
  {\bibfnamefont {R.}~\bibnamefont {Babbush}},\ }\bibfield  {title} {\enquote
  {\bibinfo {title} {Fault-tolerant quantum simulations of chemistry in first
  quantization},}\ }\href@noop {} {\bibfield  {journal} {\bibinfo  {journal}
  {PRX Quantum}\ }\textbf {\bibinfo {volume} {2}},\ \bibinfo {pages} {040332}
  (\bibinfo {year} {2021})}\BibitemShut {NoStop}%
\bibitem [{\citenamefont {Berry}\ \emph {et~al.}(2001)\citenamefont {Berry},
  \citenamefont {Wiseman},\ and\ \citenamefont {Breslin}}]{berry_optimal_2001}%
  \BibitemOpen
  \bibfield  {author} {\bibinfo {author} {\bibfnamefont {D.~W.}\ \bibnamefont
  {Berry}}, \bibinfo {author} {\bibfnamefont {H.~M.}\ \bibnamefont {Wiseman}},
  \ and\ \bibinfo {author} {\bibfnamefont {J.~K.}\ \bibnamefont {Breslin}},\
  }\bibfield  {title} {\enquote {\bibinfo {title} {Optimal input states and
  feedback for interferometric phase estimation},}\ }\href {\doibase
  10.1103/PhysRevA.63.053804} {\bibfield  {journal} {\bibinfo  {journal}
  {Physical Review A}\ }\textbf {\bibinfo {volume} {63}},\ \bibinfo {pages}
  {053804} (\bibinfo {year} {2001})}\BibitemShut {NoStop}%
\bibitem [{\citenamefont {Hentschel}\ and\ \citenamefont
  {Sanders}(2010)}]{hentschel_machine_2010}%
  \BibitemOpen
  \bibfield  {author} {\bibinfo {author} {\bibfnamefont {A.}~\bibnamefont
  {Hentschel}}\ and\ \bibinfo {author} {\bibfnamefont {B.~C.}\ \bibnamefont
  {Sanders}},\ }\bibfield  {title} {\enquote {\bibinfo {title} {Machine
  {{Learning}} for {{Precise Quantum Measurement}}},}\ }\href {\doibase
  10.1103/PhysRevLett.104.063603} {\bibfield  {journal} {\bibinfo  {journal}
  {Physical Review Letters}\ }\textbf {\bibinfo {volume} {104}},\ \bibinfo
  {pages} {063603} (\bibinfo {year} {2010})}\BibitemShut {NoStop}%
\bibitem [{\citenamefont {Wiebe}\ \emph {et~al.}(2014)\citenamefont {Wiebe},
  \citenamefont {Granade}, \citenamefont {Ferrie},\ and\ \citenamefont
  {Cory}}]{wiebe_hamiltonian_2014}%
  \BibitemOpen
  \bibfield  {author} {\bibinfo {author} {\bibfnamefont {N.}~\bibnamefont
  {Wiebe}}, \bibinfo {author} {\bibfnamefont {C.}~\bibnamefont {Granade}},
  \bibinfo {author} {\bibfnamefont {C.}~\bibnamefont {Ferrie}}, \ and\ \bibinfo
  {author} {\bibfnamefont {D.~G.}\ \bibnamefont {Cory}},\ }\bibfield  {title}
  {\enquote {\bibinfo {title} {Hamiltonian learning and certification using
  quantum resources},}\ }\href {\doibase 10.1103/PhysRevLett.112.190501}
  {\bibfield  {journal} {\bibinfo  {journal} {Physical Review Letters}\
  }\textbf {\bibinfo {volume} {112}},\ \bibinfo {pages} {190501} (\bibinfo
  {year} {2014})}\BibitemShut {NoStop}%
\bibitem [{\citenamefont {Zintchenko}\ and\ \citenamefont
  {Wiebe}(2016)}]{zintchenko_randomized_2016}%
  \BibitemOpen
  \bibfield  {author} {\bibinfo {author} {\bibfnamefont {I.}~\bibnamefont
  {Zintchenko}}\ and\ \bibinfo {author} {\bibfnamefont {N.}~\bibnamefont
  {Wiebe}},\ }\bibfield  {title} {\enquote {\bibinfo {title} {Randomized gap
  and amplitude estimation},}\ }\href {\doibase 10.1103/PhysRevA.93.062306}
  {\bibfield  {journal} {\bibinfo  {journal} {Physical Review A}\ }\textbf
  {\bibinfo {volume} {93}},\ \bibinfo {pages} {062306} (\bibinfo {year}
  {2016})}\BibitemShut {NoStop}%
\bibitem [{\citenamefont {Wang}\ \emph {et~al.}(2017)\citenamefont {Wang},
  \citenamefont {Paesani}, \citenamefont {Santagati}, \citenamefont {Knauer},
  \citenamefont {Gentile}, \citenamefont {Wiebe}, \citenamefont {Petruzzella},
  \citenamefont {O’Brien}, \citenamefont {Rarity}, \citenamefont {Laing}
  \emph {et~al.}}]{wang2017experimental}%
  \BibitemOpen
  \bibfield  {author} {\bibinfo {author} {\bibfnamefont {J.}~\bibnamefont
  {Wang}}, \bibinfo {author} {\bibfnamefont {S.}~\bibnamefont {Paesani}},
  \bibinfo {author} {\bibfnamefont {R.}~\bibnamefont {Santagati}}, \bibinfo
  {author} {\bibfnamefont {S.}~\bibnamefont {Knauer}}, \bibinfo {author}
  {\bibfnamefont {A.~A.}\ \bibnamefont {Gentile}}, \bibinfo {author}
  {\bibfnamefont {N.}~\bibnamefont {Wiebe}}, \bibinfo {author} {\bibfnamefont
  {M.}~\bibnamefont {Petruzzella}}, \bibinfo {author} {\bibfnamefont {J.~L.}\
  \bibnamefont {O’Brien}}, \bibinfo {author} {\bibfnamefont {J.~G.}\
  \bibnamefont {Rarity}}, \bibinfo {author} {\bibfnamefont {A.}~\bibnamefont
  {Laing}},  \emph {et~al.},\ }\bibfield  {title} {\enquote {\bibinfo {title}
  {Experimental quantum hamiltonian learning},}\ }\href@noop {} {\bibfield
  {journal} {\bibinfo  {journal} {Nature Physics}\ }\textbf {\bibinfo {volume}
  {13}},\ \bibinfo {pages} {551} (\bibinfo {year} {2017})}\BibitemShut
  {NoStop}%
\bibitem [{\citenamefont {Kitaev}\ \emph {et~al.}(2002)\citenamefont {Kitaev},
  \citenamefont {Shen}, \citenamefont {Vyalyi},\ and\ \citenamefont
  {Vyalyi}}]{kitaev2002classical}%
  \BibitemOpen
  \bibfield  {author} {\bibinfo {author} {\bibfnamefont {A.~Y.}\ \bibnamefont
  {Kitaev}}, \bibinfo {author} {\bibfnamefont {A.}~\bibnamefont {Shen}},
  \bibinfo {author} {\bibfnamefont {M.~N.}\ \bibnamefont {Vyalyi}}, \ and\
  \bibinfo {author} {\bibfnamefont {M.~N.}\ \bibnamefont {Vyalyi}},\
  }\href@noop {} {\emph {\bibinfo {title} {Classical and quantum
  computation}}},\ \bibinfo {number} {47}\ (\bibinfo  {publisher} {American
  Mathematical Soc.},\ \bibinfo {year} {2002})\BibitemShut {NoStop}%
\bibitem [{\citenamefont {Svore}\ \emph {et~al.}(2013)\citenamefont {Svore},
  \citenamefont {Hastings},\ and\ \citenamefont
  {Freedman}}]{svore_faster_2013}%
  \BibitemOpen
  \bibfield  {author} {\bibinfo {author} {\bibfnamefont {K.~M.}\ \bibnamefont
  {Svore}}, \bibinfo {author} {\bibfnamefont {M.~B.}\ \bibnamefont {Hastings}},
  \ and\ \bibinfo {author} {\bibfnamefont {M.}~\bibnamefont {Freedman}},\
  }\bibfield  {title} {\enquote {\bibinfo {title} {Faster phase estimation},}\
  }\href {http://arxiv.org/abs/1304.0741} {\bibfield  {journal} {\bibinfo
  {journal} {arXiv:1304.0741}\ } (\bibinfo {year} {2013})}\BibitemShut
  {NoStop}%
\bibitem [{\citenamefont {Kimmel}\ \emph {et~al.}(2015)\citenamefont {Kimmel},
  \citenamefont {Low},\ and\ \citenamefont {Yoder}}]{kimmel2015robust}%
  \BibitemOpen
  \bibfield  {author} {\bibinfo {author} {\bibfnamefont {S.}~\bibnamefont
  {Kimmel}}, \bibinfo {author} {\bibfnamefont {G.~H.}\ \bibnamefont {Low}}, \
  and\ \bibinfo {author} {\bibfnamefont {T.~J.}\ \bibnamefont {Yoder}},\
  }\bibfield  {title} {\enquote {\bibinfo {title} {Robust calibration of a
  universal single-qubit gate set via robust phase estimation},}\ }\href@noop
  {} {\bibfield  {journal} {\bibinfo  {journal} {Physical Review A}\ }\textbf
  {\bibinfo {volume} {92}},\ \bibinfo {pages} {062315} (\bibinfo {year}
  {2015})}\BibitemShut {NoStop}%
\bibitem [{\citenamefont {Wiebe}\ and\ \citenamefont
  {Granade}(2016)}]{wiebe_efficient_2015}%
  \BibitemOpen
  \bibfield  {author} {\bibinfo {author} {\bibfnamefont {N.}~\bibnamefont
  {Wiebe}}\ and\ \bibinfo {author} {\bibfnamefont {C.}~\bibnamefont
  {Granade}},\ }\bibfield  {title} {\enquote {\bibinfo {title} {Efficient
  {B}ayesian phase estimation},}\ }\href {\doibase
  10.1103/PhysRevLett.117.010503} {\bibfield  {journal} {\bibinfo  {journal}
  {Phys. Rev. Lett.}\ }\textbf {\bibinfo {volume} {117}},\ \bibinfo {pages}
  {010503} (\bibinfo {year} {2016})}\BibitemShut {NoStop}%
\bibitem [{\citenamefont {Paesani}\ \emph {et~al.}(2017)\citenamefont
  {Paesani}, \citenamefont {Gentile}, \citenamefont {Santagati}, \citenamefont
  {Wang}, \citenamefont {Wiebe}, \citenamefont {Tew}, \citenamefont {O'Brien},\
  and\ \citenamefont {Thompson}}]{paesani_experimental_2017}%
  \BibitemOpen
  \bibfield  {author} {\bibinfo {author} {\bibfnamefont {S.}~\bibnamefont
  {Paesani}}, \bibinfo {author} {\bibfnamefont {A.~A.}\ \bibnamefont
  {Gentile}}, \bibinfo {author} {\bibfnamefont {R.}~\bibnamefont {Santagati}},
  \bibinfo {author} {\bibfnamefont {J.}~\bibnamefont {Wang}}, \bibinfo {author}
  {\bibfnamefont {N.}~\bibnamefont {Wiebe}}, \bibinfo {author} {\bibfnamefont
  {D.~P.}\ \bibnamefont {Tew}}, \bibinfo {author} {\bibfnamefont {J.~L.}\
  \bibnamefont {O'Brien}}, \ and\ \bibinfo {author} {\bibfnamefont {M.~G.}\
  \bibnamefont {Thompson}},\ }\bibfield  {title} {\enquote {\bibinfo {title}
  {Experimental {{Bayesian Quantum Phase Estimation}} on a {{Silicon Photonic
  Chip}}},}\ }\href {\doibase 10.1103/PhysRevLett.118.100503} {\bibfield
  {journal} {\bibinfo  {journal} {Physical Review Letters}\ }\textbf {\bibinfo
  {volume} {118}},\ \bibinfo {pages} {100503} (\bibinfo {year}
  {2017})}\BibitemShut {NoStop}%
\bibitem [{\citenamefont {Eldredge}\ \emph {et~al.}(2016)\citenamefont
  {Eldredge}, \citenamefont {Foss-Feig}, \citenamefont {Rolston},\ and\
  \citenamefont {Gorshkov}}]{eldredge_optimal_2016}%
  \BibitemOpen
  \bibfield  {author} {\bibinfo {author} {\bibfnamefont {Z.}~\bibnamefont
  {Eldredge}}, \bibinfo {author} {\bibfnamefont {M.}~\bibnamefont {Foss-Feig}},
  \bibinfo {author} {\bibfnamefont {S.~L.}\ \bibnamefont {Rolston}}, \ and\
  \bibinfo {author} {\bibfnamefont {A.~V.}\ \bibnamefont {Gorshkov}},\
  }\bibfield  {title} {\enquote {\bibinfo {title} {Optimal and secure
  measurement protocols for quantum sensor networks},}\ }\href
  {http://arxiv.org/abs/1607.04646} {\bibfield  {journal} {\bibinfo  {journal}
  {arXiv:1607.04646 [quant-ph]}\ } (\bibinfo {year} {2016})},\ \bibinfo {note}
  {arXiv: 1607.04646}\BibitemShut {NoStop}%
\bibitem [{\citenamefont
  {Kitaev}(1995)}]{kitaevay_quantummeasurementsabelian_1995}%
  \BibitemOpen
  \bibfield  {author} {\bibinfo {author} {\bibfnamefont {A.~Y.}\ \bibnamefont
  {Kitaev}},\ }\bibfield  {title} {\enquote {\bibinfo {title} {Quantum
  measurements and the {{Abelian}} stabilizer problem},}\ }\href@noop {}
  {\bibfield  {journal} {\bibinfo  {journal} {arXiv:quant-ph/9511026}\ }
  (\bibinfo {year} {1995})},\ \Eprint {http://arxiv.org/abs/quant-ph/9511026}
  {quant-ph/9511026} \BibitemShut {NoStop}%
\bibitem [{\citenamefont {Gily{\'e}n}\ \emph {et~al.}(2019)\citenamefont
  {Gily{\'e}n}, \citenamefont {Su}, \citenamefont {Low},\ and\ \citenamefont
  {Wiebe}}]{gilyen2019quantum}%
  \BibitemOpen
  \bibfield  {author} {\bibinfo {author} {\bibfnamefont {A.}~\bibnamefont
  {Gily{\'e}n}}, \bibinfo {author} {\bibfnamefont {Y.}~\bibnamefont {Su}},
  \bibinfo {author} {\bibfnamefont {G.~H.}\ \bibnamefont {Low}}, \ and\
  \bibinfo {author} {\bibfnamefont {N.}~\bibnamefont {Wiebe}},\ }\bibfield
  {title} {\enquote {\bibinfo {title} {Quantum singular value transformation
  and beyond: exponential improvements for quantum matrix arithmetics},}\ }in\
  \href@noop {} {\emph {\bibinfo {booktitle} {Proceedings of the 51st Annual
  ACM SIGACT Symposium on Theory of Computing}}}\ (\bibinfo {year} {2019})\
  pp.\ \bibinfo {pages} {193--204}\BibitemShut {NoStop}%
\bibitem [{\citenamefont {Wiebe}\ \emph {et~al.}(2015)\citenamefont {Wiebe},
  \citenamefont {Granade},\ and\ \citenamefont
  {Cory}}]{wiebeQuantumBootstrappingCompressed2015}%
  \BibitemOpen
  \bibfield  {author} {\bibinfo {author} {\bibfnamefont {N.}~\bibnamefont
  {Wiebe}}, \bibinfo {author} {\bibfnamefont {C.}~\bibnamefont {Granade}}, \
  and\ \bibinfo {author} {\bibfnamefont {D.~G.}\ \bibnamefont {Cory}},\
  }\bibfield  {title} {\enquote {\bibinfo {title} {Quantum bootstrapping via
  compressed quantum {{Hamiltonian}} learning},}\ }\href {\doibase
  10.1088/1367-2630/17/2/022005} {\bibfield  {journal} {\bibinfo  {journal}
  {New Journal of Physics}\ }\textbf {\bibinfo {volume} {17}},\ \bibinfo
  {pages} {022005} (\bibinfo {year} {2015})}\BibitemShut {NoStop}%
\bibitem [{\citenamefont {Tipireddy}\ and\ \citenamefont
  {Wiebe}(2020)}]{tipireddy2020bayesian}%
  \BibitemOpen
  \bibfield  {author} {\bibinfo {author} {\bibfnamefont {R.}~\bibnamefont
  {Tipireddy}}\ and\ \bibinfo {author} {\bibfnamefont {N.}~\bibnamefont
  {Wiebe}},\ }\bibfield  {title} {\enquote {\bibinfo {title} {Bayesian phase
  estimation with adaptive grid refinement},}\ }\href@noop {} {\bibfield
  {journal} {\bibinfo  {journal} {arXiv preprint arXiv:2009.07898}\ } (\bibinfo
  {year} {2020})}\BibitemShut {NoStop}%
\bibitem [{\citenamefont {Lumino}\ \emph {et~al.}(2018)\citenamefont {Lumino},
  \citenamefont {Polino}, \citenamefont {Rab}, \citenamefont {Milani},
  \citenamefont {Spagnolo}, \citenamefont {Wiebe},\ and\ \citenamefont
  {Sciarrino}}]{lumino2018experimental}%
  \BibitemOpen
  \bibfield  {author} {\bibinfo {author} {\bibfnamefont {A.}~\bibnamefont
  {Lumino}}, \bibinfo {author} {\bibfnamefont {E.}~\bibnamefont {Polino}},
  \bibinfo {author} {\bibfnamefont {A.~S.}\ \bibnamefont {Rab}}, \bibinfo
  {author} {\bibfnamefont {G.}~\bibnamefont {Milani}}, \bibinfo {author}
  {\bibfnamefont {N.}~\bibnamefont {Spagnolo}}, \bibinfo {author}
  {\bibfnamefont {N.}~\bibnamefont {Wiebe}}, \ and\ \bibinfo {author}
  {\bibfnamefont {F.}~\bibnamefont {Sciarrino}},\ }\bibfield  {title} {\enquote
  {\bibinfo {title} {Experimental phase estimation enhanced by machine
  learning},}\ }\href@noop {} {\bibfield  {journal} {\bibinfo  {journal}
  {Physical Review Applied}\ }\textbf {\bibinfo {volume} {10}},\ \bibinfo
  {pages} {044033} (\bibinfo {year} {2018})}\BibitemShut {NoStop}%
\bibitem [{\citenamefont {Doucet}\ and\ \citenamefont
  {Johansen}(2011)}]{doucet_tutorial_2011}%
  \BibitemOpen
  \bibfield  {author} {\bibinfo {author} {\bibfnamefont {A.}~\bibnamefont
  {Doucet}}\ and\ \bibinfo {author} {\bibfnamefont {A.~M.}\ \bibnamefont
  {Johansen}},\ }\href@noop {} {\emph {\bibinfo {title} {A Tutorial on Particle
  Filtering and Smoothing: Fifteen Years Later}}}\ (\bibinfo {year}
  {2011})\BibitemShut {NoStop}%
\bibitem [{\citenamefont {Granade}\ \emph {et~al.}(2017)\citenamefont
  {Granade}, \citenamefont {Ferrie}, \citenamefont {Hincks}, \citenamefont
  {Casagrande}, \citenamefont {Alexander}, \citenamefont {Gross}, \citenamefont
  {Kononenko},\ and\ \citenamefont {Sanders}}]{granade_qinfer_2017}%
  \BibitemOpen
  \bibfield  {author} {\bibinfo {author} {\bibfnamefont {C.}~\bibnamefont
  {Granade}}, \bibinfo {author} {\bibfnamefont {C.}~\bibnamefont {Ferrie}},
  \bibinfo {author} {\bibfnamefont {I.}~\bibnamefont {Hincks}}, \bibinfo
  {author} {\bibfnamefont {S.}~\bibnamefont {Casagrande}}, \bibinfo {author}
  {\bibfnamefont {T.}~\bibnamefont {Alexander}}, \bibinfo {author}
  {\bibfnamefont {J.}~\bibnamefont {Gross}}, \bibinfo {author} {\bibfnamefont
  {M.}~\bibnamefont {Kononenko}}, \ and\ \bibinfo {author} {\bibfnamefont
  {Y.}~\bibnamefont {Sanders}},\ }\bibfield  {title} {\enquote {\bibinfo
  {title} {{QI}nfer: {S}tatistical inference software for quantum
  applications},}\ }\href {\doibase 10.22331/q-2017-04-25-5} {\bibfield
  {journal} {\bibinfo  {journal} {{Quantum}}\ }\textbf {\bibinfo {volume}
  {1}},\ \bibinfo {pages} {5} (\bibinfo {year} {2017})}\BibitemShut {NoStop}%
\bibitem [{\citenamefont {Banerjee}\ \emph {et~al.}(2005)\citenamefont
  {Banerjee}, \citenamefont {Guo},\ and\ \citenamefont
  {Wang}}]{banerjeeOptimalityConditionalExpectation2005}%
  \BibitemOpen
  \bibfield  {author} {\bibinfo {author} {\bibfnamefont {A.}~\bibnamefont
  {Banerjee}}, \bibinfo {author} {\bibfnamefont {X.}~\bibnamefont {Guo}}, \
  and\ \bibinfo {author} {\bibfnamefont {H.}~\bibnamefont {Wang}},\ }\bibfield
  {title} {\enquote {\bibinfo {title} {On the optimality of conditional
  expectation as a {{Bregman}} predictor},}\ }\href {\doibase
  10.1109/TIT.2005.850145} {\bibfield  {journal} {\bibinfo  {journal} {IEEE
  Transactions on Information Theory}\ }\textbf {\bibinfo {volume} {51}},\
  \bibinfo {pages} {2664} (\bibinfo {year} {2005})}\BibitemShut {NoStop}%
\bibitem [{\citenamefont {Gill}\ and\ \citenamefont
  {Levit}(1995)}]{gillApplicationsVanTrees1995}%
  \BibitemOpen
  \bibfield  {author} {\bibinfo {author} {\bibfnamefont {R.~D.}\ \bibnamefont
  {Gill}}\ and\ \bibinfo {author} {\bibfnamefont {B.~Y.}\ \bibnamefont
  {Levit}},\ }\bibfield  {title} {\enquote {\bibinfo {title} {Applications of
  the van {{Trees}} inequality: A {{Bayesian Cram\'er-Rao}} bound},}\
  }\href@noop {} {\bibfield  {journal} {\bibinfo  {journal} {Bernoulli}\
  }\textbf {\bibinfo {volume} {1}},\ \bibinfo {pages} {59} (\bibinfo {year}
  {1995})}\BibitemShut {NoStop}%
\bibitem [{\citenamefont {Cover}\ and\ \citenamefont
  {Thomas}(1991)}]{cover1991information}%
  \BibitemOpen
  \bibfield  {author} {\bibinfo {author} {\bibfnamefont {T.~M.}\ \bibnamefont
  {Cover}}\ and\ \bibinfo {author} {\bibfnamefont {J.~A.}\ \bibnamefont
  {Thomas}},\ }\bibfield  {title} {\enquote {\bibinfo {title} {Information
  theory and statistics},}\ }\href@noop {} {\bibfield  {journal} {\bibinfo
  {journal} {Elements of information theory}\ }\textbf {\bibinfo {volume}
  {1}},\ \bibinfo {pages} {279} (\bibinfo {year} {1991})}\BibitemShut {NoStop}%
\bibitem [{\citenamefont {Opper}(1998)}]{Opper:1999:BAO:304710.304756}%
  \BibitemOpen
  \bibfield  {author} {\bibinfo {author} {\bibfnamefont {M.}~\bibnamefont
  {Opper}},\ }\bibfield  {title} {\enquote {\bibinfo {title} {On-line learning
  in neural networks},}\ \ }(\bibinfo  {publisher} {{Cambridge University
  Press}},\ \bibinfo {address} {{New York, NY, USA}},\ \bibinfo {year} {1998})\
  pp.\ \bibinfo {pages} {363--378}\BibitemShut {NoStop}%
\bibitem [{\citenamefont {Ferrie}\ \emph {et~al.}(2013)\citenamefont {Ferrie},
  \citenamefont {Granade},\ and\ \citenamefont
  {Cory}}]{ferrieHowBestSample2013}%
  \BibitemOpen
  \bibfield  {author} {\bibinfo {author} {\bibfnamefont {C.}~\bibnamefont
  {Ferrie}}, \bibinfo {author} {\bibfnamefont {C.~E.}\ \bibnamefont {Granade}},
  \ and\ \bibinfo {author} {\bibfnamefont {D.~G.}\ \bibnamefont {Cory}},\
  }\bibfield  {title} {\enquote {\bibinfo {title} {How to best sample a
  periodic probability distribution, or on the accuracy of {{Hamiltonian}}
  finding strategies},}\ }\href {\doibase 10.1007/s11128-012-0407-6} {\bibfield
   {journal} {\bibinfo  {journal} {Quantum Information Processing}\ }\textbf
  {\bibinfo {volume} {12}},\ \bibinfo {pages} {611} (\bibinfo {year}
  {2013})}\BibitemShut {NoStop}%
\bibitem [{\citenamefont {Combes}\ \emph {et~al.}(2014)\citenamefont {Combes},
  \citenamefont {Ferrie}, \citenamefont {Jiang},\ and\ \citenamefont
  {Caves}}]{combes2014quantum}%
  \BibitemOpen
  \bibfield  {author} {\bibinfo {author} {\bibfnamefont {J.}~\bibnamefont
  {Combes}}, \bibinfo {author} {\bibfnamefont {C.}~\bibnamefont {Ferrie}},
  \bibinfo {author} {\bibfnamefont {Z.}~\bibnamefont {Jiang}}, \ and\ \bibinfo
  {author} {\bibfnamefont {C.~M.}\ \bibnamefont {Caves}},\ }\bibfield  {title}
  {\enquote {\bibinfo {title} {Quantum limits on postselected, probabilistic
  quantum metrology},}\ }\href@noop {} {\bibfield  {journal} {\bibinfo
  {journal} {Physical Review A}\ }\textbf {\bibinfo {volume} {89}},\ \bibinfo
  {pages} {052117} (\bibinfo {year} {2014})}\BibitemShut {NoStop}%
\bibitem [{\citenamefont {Granade}\ \emph {et~al.}(2012)\citenamefont
  {Granade}, \citenamefont {Ferrie}, \citenamefont {Wiebe},\ and\ \citenamefont
  {Cory}}]{granade2012robust}%
  \BibitemOpen
  \bibfield  {author} {\bibinfo {author} {\bibfnamefont {C.~E.}\ \bibnamefont
  {Granade}}, \bibinfo {author} {\bibfnamefont {C.}~\bibnamefont {Ferrie}},
  \bibinfo {author} {\bibfnamefont {N.}~\bibnamefont {Wiebe}}, \ and\ \bibinfo
  {author} {\bibfnamefont {D.~G.}\ \bibnamefont {Cory}},\ }\bibfield  {title}
  {\enquote {\bibinfo {title} {Robust online hamiltonian learning},}\
  }\href@noop {} {\bibfield  {journal} {\bibinfo  {journal} {New Journal of
  Physics}\ }\textbf {\bibinfo {volume} {14}},\ \bibinfo {pages} {103013}
  (\bibinfo {year} {2012})}\BibitemShut {NoStop}%
\end{thebibliography}%

\begin{widetext}
\begin{minipage}{\linewidth}
\begin{algorithm}[H]
    \caption{\label{alg:rwpe-w-unwinding}
        Random walk phase estimation algorithm with
        consistency checks and data unwinding.
    }
    \begin{algorithmic}
        
        \Function{RandomWalkPhaseEst}{
                $\mu_0$, $\sigma_0$,
                $\tch$, $n_\unwind$
        }
            \Arguments
                \State $\mu_0$: initial mean
                \State $\sigma_0$: initial standard deviation
                \State $\tch$: consistency check scale
                \State $n_\unwind$: number of unwinding steps
            \EndArguments
            \seccomment{Initialization}
            \State $\mu \gets \mu_0$
            \State $\sigma \gets \sigma_0$
            \State $D \gets$ a stack of Boolean values.
            \seccomment{Main body}
            \For{$\iexp \in \{0, 1, \dots \nexp - 1\}$}
                \State $\omega_{\inv} \gets \mu - \pi \sigma / 2$
                \State $t \gets 1 / \sigma$
                \State Sample a datum $d$ from $\Pr(d = 0 | \omega; \omega_\inv, t) = \cos^2(t (\omega - \omega_\inv) / 2)$.
                \State Push $d$ onto $D$.
                \If{$d = 0$}
                    \State $\mu \gets \mu + \sigma / \sqrt{e}$
                \Else
                    \State $\mu \gets \mu - \sigma / \sqrt{e}$
                \EndIf
                \State $\sigma \gets \sigma \sqrt{(e - 1) / e}$

                \seccomment{Data unwinding}
                \If{$n_\unwind > 0$}

                \linecomment{Perform a consistency check; $d' = 0$ is most probable if our approximation is correct.}
                \State Sample a datum $d'$ from $\Pr(d' = 0 | \omega; \mu, \tch / \sigma) = \cos^2(\tch / \sigma (\omega - \mu) / 2)$.
                \linecomment{Keep unwinding until the consistency check passes.}
                \While{$d' = 1$}
                    \For{$i_\unwind \in \{0, \dots, n_\unwind - 1\}$}
                        \State $\sigma \gets \sigma \sqrt{e / (e  - 1)}$
                        \If {$D$ is not empty}
                            \inlinecomment{Only shift $\mu$ backwards when we unwound actual data.}
                            \State $d \gets \operatorname{pop} D$
                            \If{$d = 0$}
                                \State $\mu \gets \mu - \sigma / \sqrt{e}$
                            \Else
                                \State $\mu \gets \mu + \sigma / \sqrt{e}$
                            \EndIf
                        \EndIf
                    \EndFor
                    \linecomment{Perform a new consistency check.}
                    \State Sample a datum $d'$ from $\Pr(d' = 0 | \omega; \mu, \tch / \sigma) = \cos^2(\tch / \sigma (\omega - \mu) / 2)$.
                \EndWhile
                
                \EndIf

            \EndFor
            \seccomment{Final estimate}
            \State \Return $\hat{\omega} \gets \mu$
        \EndFunction
    \end{algorithmic}
\end{algorithm}
\end{minipage}
\end{widetext}

\appendix

\section{Unwinding Past the Initial Prior}
\newpage
In \autoref{fig:alg-loss-histogram}, we presented the results of using \autoref{alg:rwpe-w-unwinding} with unwinding steps that can continue backwards past the initial prior. Effectively, unwinding past the initial prior automates interventions such as those used by rejection filtering PE \cite{wiebe_efficient_2015} by allowing RWPE to explore hypothesis that are considered improbable under the initial prior distribution.

\begin{figure*}
    \begin{center}
        \includegraphics[width=0.9\textwidth]{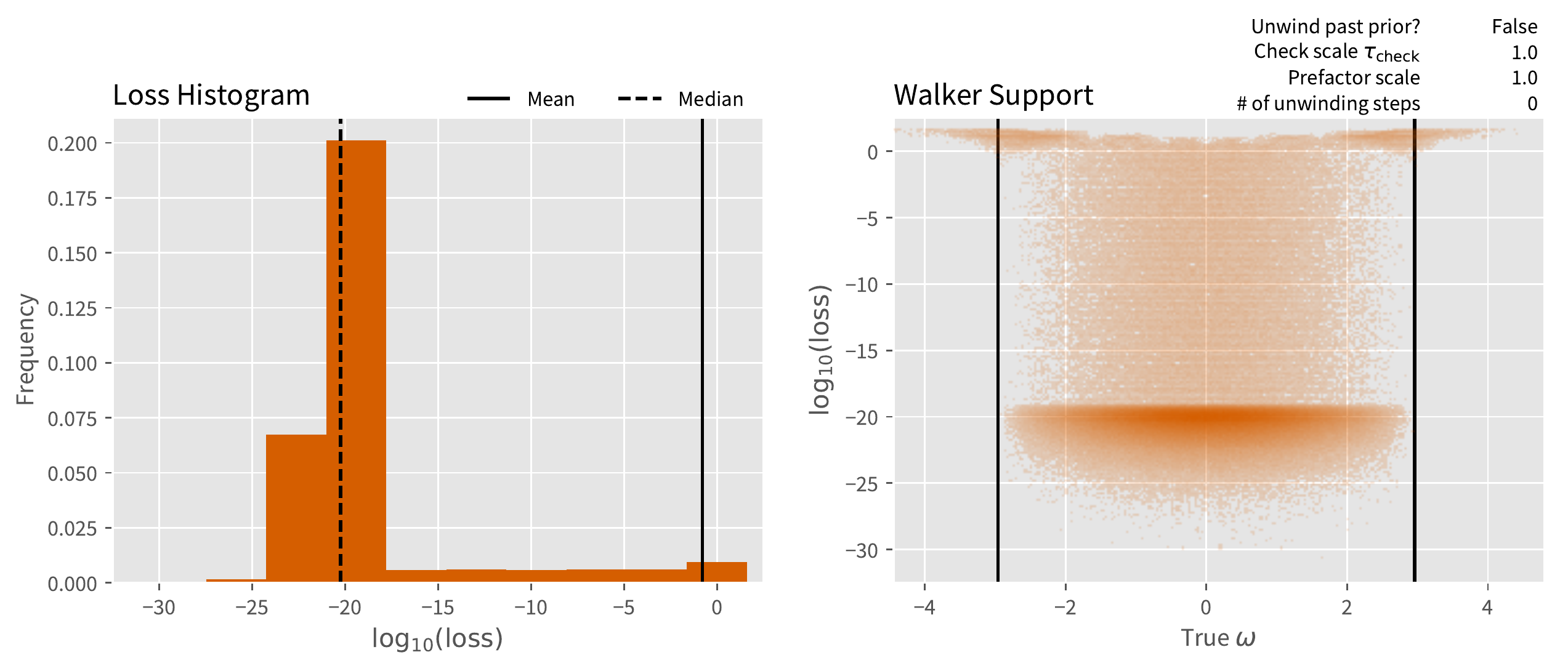}

        \includegraphics[width=0.9\textwidth]{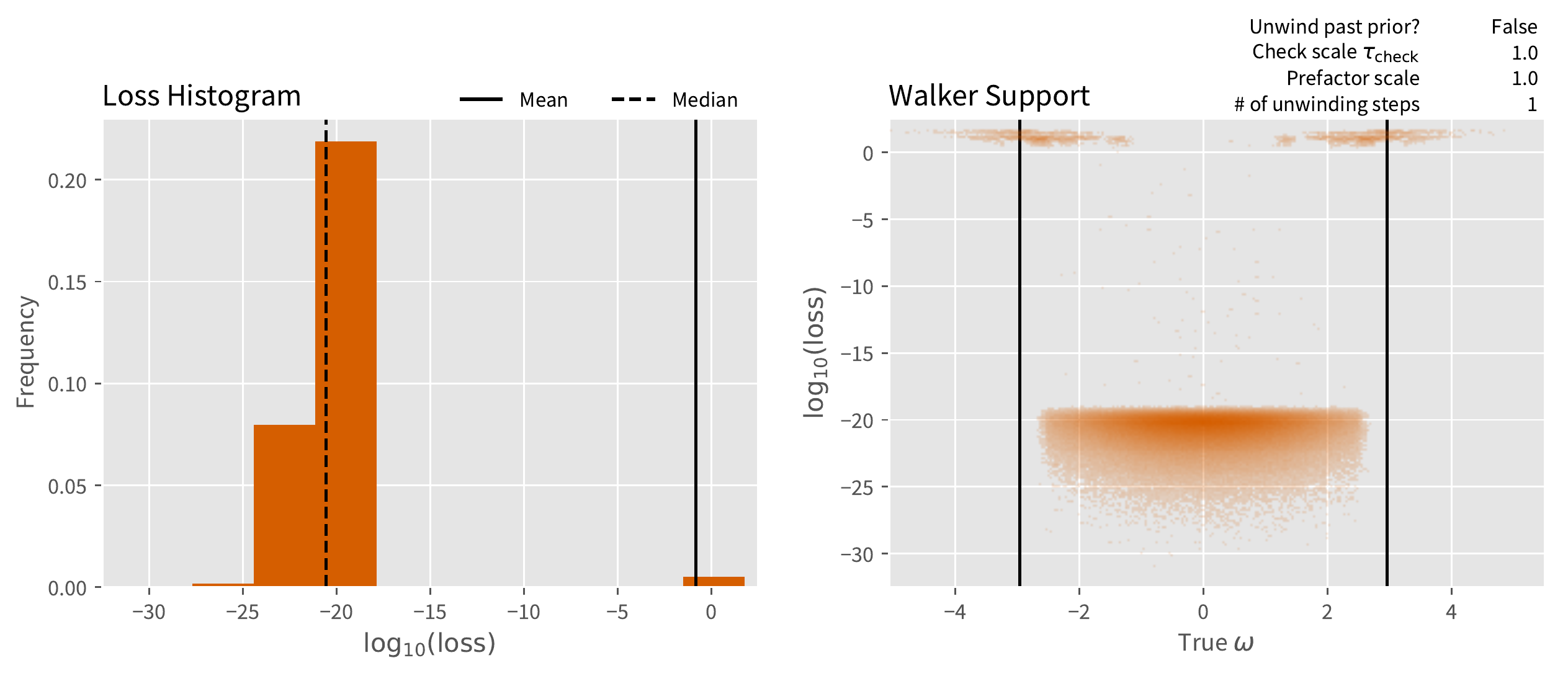}

        \includegraphics[width=0.9\textwidth]{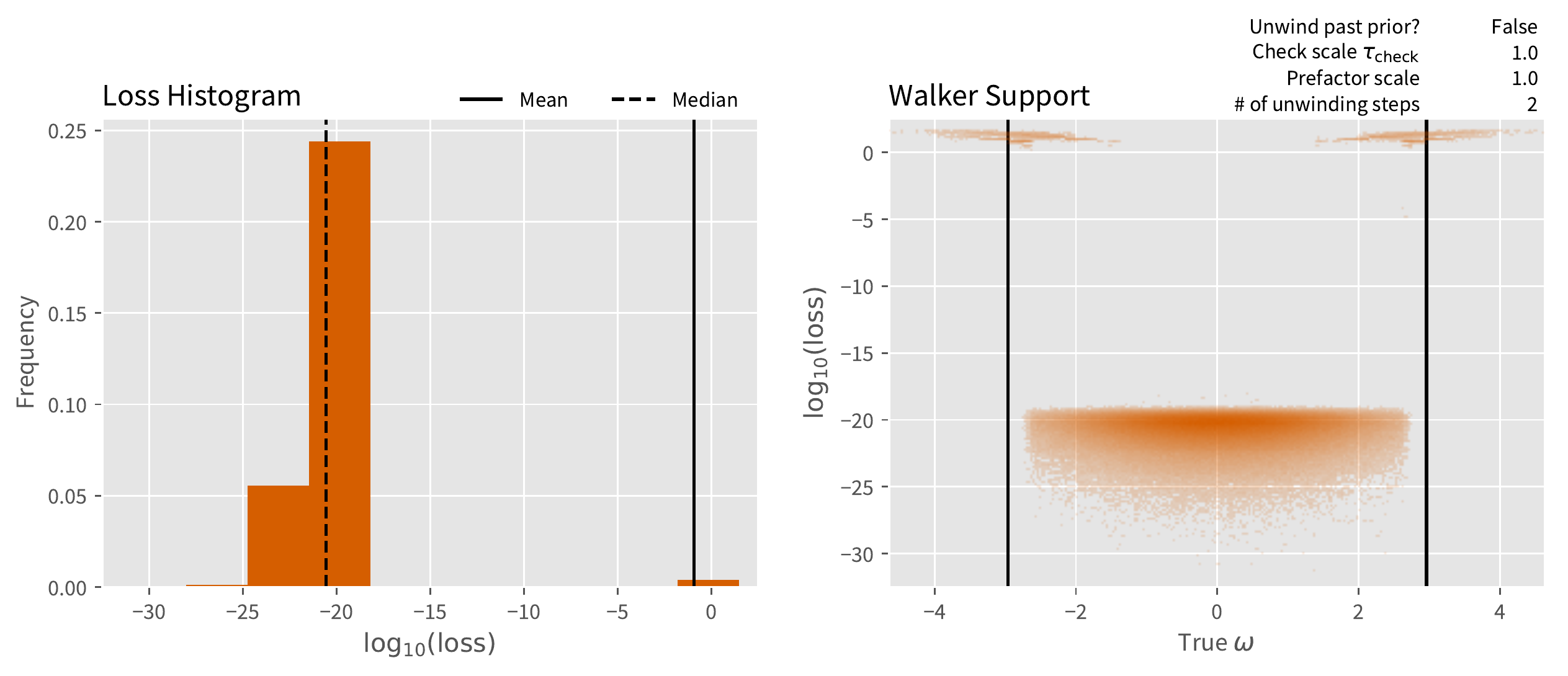}
    \end{center}
    \caption{
        \label{fig:alg-loss-histogram-no-neg}
        \textbf{(Left column)}
            Histogram over log-losses for many different trials, compared to the mean loss (Bayes risk)
            and the median loss.
        \textbf{(Right column)}
            Log-loss versus the true value of $\omega$, compared with the finite range (thick lines)
            in which each walker can explore.
        \textbf{(Top row)}
            Trials are run using the basic approach of \autoref{alg:rwpe-no-unwinding}.
        \textbf{(Middle row)}
            Trials are run using one constrained unwinding step, as in \autoref{alg:rwpe-w-unwinding}.
        \textbf{(Bottom row)}
            Trials are run using two constrained unwinding steps, as in \autoref{alg:rwpe-w-unwinding}.
    }
\end{figure*}

In this Appendix, we underscore that this rule is critical to allow the random walk to exceed the finite constraints of the basic walk presented \autoref{alg:rwpe-no-unwinding} by repeating the previous analysis with unwinding on actual data only.
That is, in the results presented within \autoref{fig:alg-loss-histogram-no-neg}, if an unwinding step would pop an empty stack $D$, then the unwinding step is aborted instead.

\begin{figure*}
    \paragraph*{Constrained Unwinding}
    \begin{center}
        \includegraphics[width=0.95\textwidth]{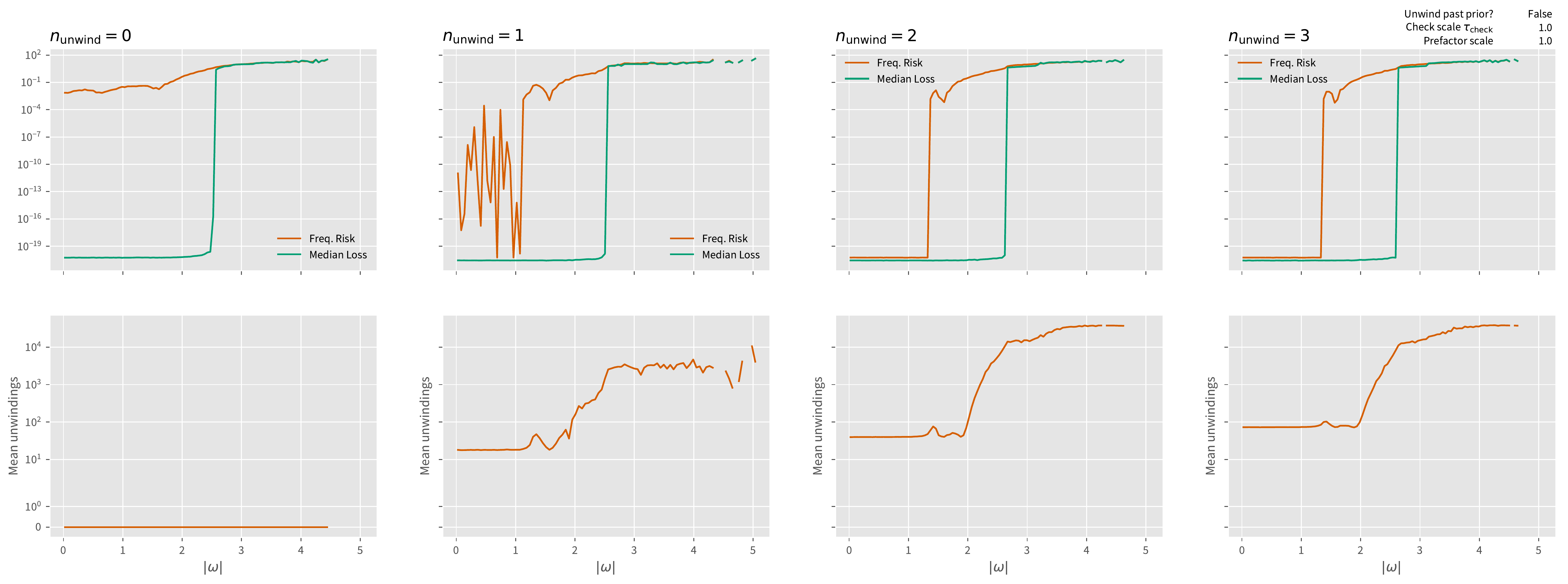}
    \end{center}
    \paragraph*{Unwinding Past Initial Prior}
    \begin{center}
        \includegraphics[width=0.95\textwidth]{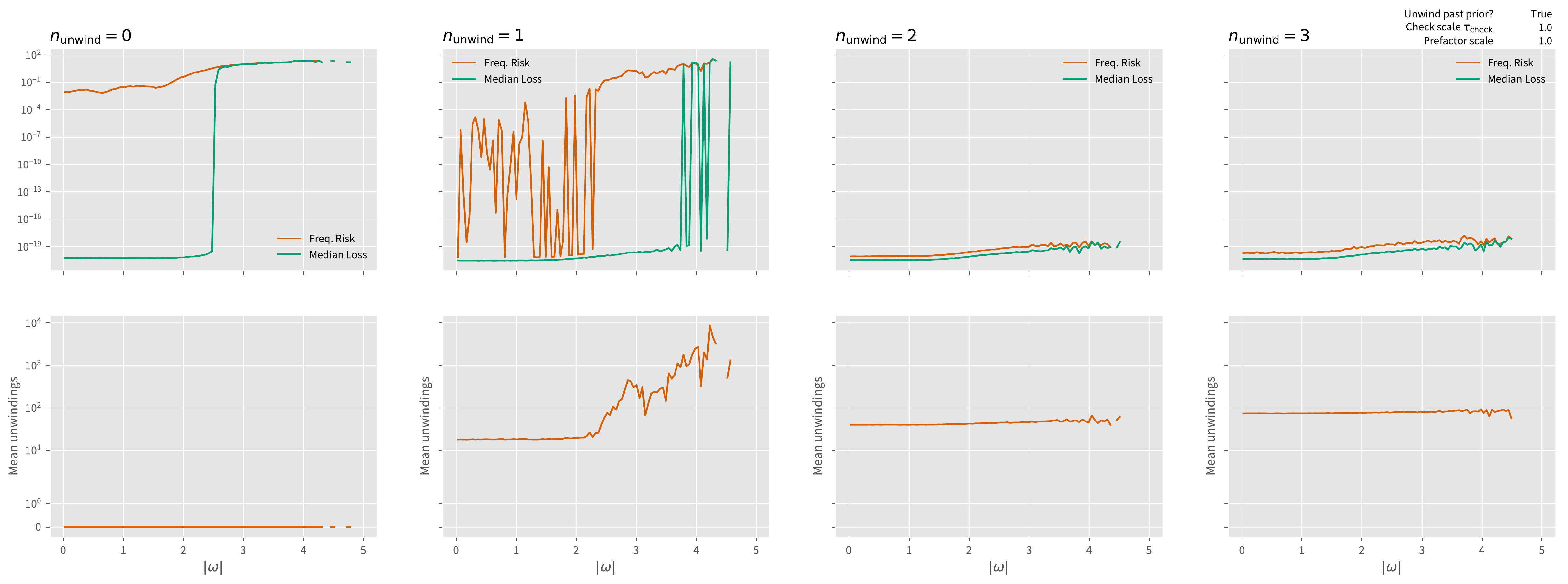}
    \end{center}
    \caption{
        \label{fig:risk-profiles}
        \textbf{(Top section)}
            Unwinding constrained by initial prior.
        \textbf{(Bottom section)}
            Unwinding as described in \autoref{alg:rwpe-w-unwinding}.
        \textbf{(Top row of each section)}
            Frequentist risk and median loss as a function of $|\omega|$, the absolute value of the true $\omega$ used to generate data in each trial.
        \textbf{(Bottom row of each section)}
            Median number of unwinding steps taken to accept 100 experiments.
        \textbf{(Columns)}
            Different numbers of unwinding steps allowed.
    }
\end{figure*}

We note that the constrained unwinding, while still an improvement over the initial version of \autoref{alg:rwpe-no-unwinding}, does not adequately deal with the effects of the finite support of the basic walk.
This observation is underscored by the results presented in \autoref{fig:risk-profiles}, where we consider the frequentist risk rather than the traditional Bayes risk.
Concretely, the frequentist risk is defined as the average loss that incurred by an estimation when the true value of $\omega$ is fixed, $R(\hat{\omega}, \omega) \defeq \expect_D[(\hat{\omega} - \omega)^2]$.
We then observe that without allowing the unwinding step to proceed backwards from the initial prior, the frequentist risk suddenly worsens by approximately $10^{20}$ when the finite extent of the random walk begins to dominate.

By contrast, the frequentist risk for RWPE with unwinding of at least two steps past the initial prior is much more flat, representing that the risk remains acceptable across the full range of valid hypotheses.



\end{document}